\documentclass[a4paper,11pt]{article}
\pdfoutput=1 % if your are submitting a pdflatex (i.e. if you have
\usepackage{jcappub} % for details on the use of the package, please
\usepackage[T1]{fontenc} % if needed

\newcommand{\bq}{\boldsymbol q}
\newcommand{\bx}{\boldsymbol x}
\newcommand{\bk}{\textbf{k}}
\newcommand{\bPsi}{\boldsymbol{\Psi}}

\newcommand{\ihmpc}{\,h{\rm Mpc}^{-1}}
\newcommand{\aemulus}{\texttt{Aemulus} }
\newcommand{\quijote}{\texttt{Quijote} }
\newcommand{\zenbu}{\texttt{ZeNBu}}

\newcommand{\monofonic}{\texttt{monofonIC} }
\newcommand{\velocileptors}{\texttt{velocileptors} }

\title{\boldmath Accurate predictions from small boxes: variance suppression via the Zel'dovich approximation}

\author[a,b]{Nickolas Kokron}
\author[c]{Shi-Fan Chen}
\author[c,d]{Martin White} 
\author[d]{Joseph DeRose}
\author[c]{Mark Maus}
\emailAdd{kokron@stanford.edu}

\affiliation[a]{Kavli Institute for Particle Astrophysics and Cosmology and Department of Physics, Stanford University, Stanford, CA, USA}
\affiliation[b]{Kavli Institute for Particle Astrophysics and Cosmology, SLAC National Accelerator Laboratory, Menlo Park, CA, USA}
\affiliation[c]{Department of Physics, University of California, Berkeley, CA, USA}
\affiliation[d]{Physics Division, Lawrence Berkeley National Laboratory, Berkeley, CA, USA}

\abstract{
Simulations have become an indispensable tool for accurate modelling of observables measured in galaxy surveys, but can be expensive if very large dynamic range in scale is required.  We describe how to combine Lagrangian perturbation theory models with N-body simulations to reduce the effects of finite computational volume in the prediction of ensemble average properties in the simulations within the context of control variates.  In particular we use the fact that Zel'dovich displacements, computed during initial condition generation for any simulation, correlate strongly with the final density field.  Since all the correlators of biased tracers can be computed with arbitrary precision for these displacements, pairing the Zel'dovich `simulation' with the N-body realization allows hundredfold reductions in sample variance for power spectrum or correlation function estimation. Zel'dovich control variates can accurately extend matter or tracer field emulators to larger scales than previously possible, as well as improving measurements of statistics in simulations which are inherently limited to small volumes, such as hydrodynamical simulations of galaxy formation and reionization.
}

\begin{document}
\maketitle
\flushbottom

\section{Introduction}
\label{sec:intro}

Computer simulations of the formation of cosmic structures, from those that include dark matter only to those solving for full radiative hydrodynamics, have become essential tools in understanding the evolution of the Universe across different cosmic eras. Simulations shed light on complicated non-linear phenomena that evade analytic descriptions. These include: the non-linear dynamics of gravitational collapse, leading to the formation of bound halos; the interplay between gas and light, leading to the formation of stars and eventually galaxies within these halos; the radiative processes underlying reionization and countless more. \par 
In numerical cosmology there exists an eternal tug-of-war between accurately resolving small-scale physics and running simulations at large enough volumes that one has a statistically robust result. For example, suites of simulations run at numerous cosmologies, used for the construction of emulators, must not only balance the dynamic range of their simulations but also include the expense of maintaining accuracy while spanning a large space of cosmological parameters, where a new simulation has to be run for each point in this parameter space.
For emulators of large-scale structure designed to accurately predict summary statistics measured in galaxy surveys, the challenges associated with limited volumes become even more severe. Next-generation galaxy surveys will probe unprecedented cosmic volumes, requiring highly accurate (and precise) predictions at large scales, imposing additional requirements on simulation-based inference tools. \par 
To surmount these challenges, a plethora of techniques have been introduced that try to ameliorate either resolution or volume requirements of simulations. On the resolution front, statistical learning techniques have been leveraged to produce algorithms that generate so-called \emph{super-resolution} simulations which ``fill in'' information on scales smaller than what the original simulation was capable of resolving in an inexpensive way \cite{Dai_2018, Dai_2020, Dai_2021, Li_2021, Schaurecker:2021lkd}. However, super-resolution techniques are still in their infancy and must be understood more before being used to confront data from these galaxy surveys. On the other hand, the field of statistics has a rich literature on the subject of variance reduction, and these techniques have begun to be imported in a cosmological context in order to relax the requirements on either the volumes of simulations or sheer quantity which must be run. Perhaps the most popular variance reduction tool in cosmology is that of Latin Hypercube Sampling, which allows for efficient sampling of high-dimensional parameter spaces such as the 7(8)-dimensional $ w (\nu)$CDM space over which emulator suites are constructed \cite{miratitan, eucludemu2,DeRose:2018xdj,Maksimova:2021ynf}. Two other techniques which have seen widespread adoption are ``paired phase'' and ``fixed amplitude'' simulations. Paired-phase, or, ``pairing'', involves simulating two Universes whose initial conditions are exactly the same up to a minus sign \cite{Pontzen_2016}. The mean of statistics computed in each simulation then has its variance significantly reduced relative to the expectation of Gaussian initial conditions. ``Fixing'', on the other hand, involves initializing simulations where the amplitude of density fluctuations follows a Dirac delta distribution as opposed to the standard Rayleigh distribution \cite{Angulo:2016hjd}. Fixing also significantly reduces the large-scale variance of an $N$-body simulation. The combination of these two techniques, ``paired--fixed'' simulations, has become the object of significant study in recent years \cite{Villaescusa-Navarro:2018bpd, Chuang:2018ega,Anderson:2018zkm, Klypin_2020}. \par 
Another technique which has recently seen use in reducing sample variance of simulations in cosmology is the method of control variates \cite{chartier2020,chartier2021,Chartier:2022kjz}. Control variates are particularly powerful when correlated, inexpensive surrogates of the statistics one wishes to measure can be produced. The method is well-understood from a theoretical point of view, and the potential variance reduction that can be achieved through its optimal application is proportional to the degree of correlation between the surrogate adopted and the costly statistic whose variance we wish to reduce. The success of control variates is predicated on a thorough understanding of the statistics of the surrogate adopted, including a well-characterized mean, variance and its co-variance with the desired statistic. So far, the surrogate of choice adopted in cosmological applications has been an approximate simulation which, while significantly less expensive than a full $N$-body simulation to produce, still incurs a substantial computational cost due to the need of simulating hundreds of approximate mocks in order to estimate the mean of the surrogate. Thus, the current bottleneck of applying control variates to cosmological simulations lies in the requirement of simulating large numbers of approximate simulations in order to characterize well its statistical properties.  \par
At sufficiently large scales, analytic descriptions of large-scale structure statistics are highly accurate and arbitrarily precise, and are thus powerful tools to study the large-scale regime of large-volume surveys. While traditionally treated as two disparate ways of studying structure formation, analytic and simulation-based descriptions are inherently linked. For example, every simulation of cosmic structure requires initial conditions which are generated from the aforementioned analytic descriptions, specifically using Lagrangian Perturbation Theory (LPT). LPT is not only a potent framework to describe the statistics of the densities and velocities of biased tracers \cite{Matsubara_2008,Matsubara2008,Carlson_2012,Porto14,Vlah15,Vlah_2016,Chen_2020,Chen_2021,Chen22a,white2022cosmological}, but it has recently also been used in combination with $N$-body simulations to produce \emph{hybrid} models of structure formation \cite{modichenwhite19}. Hybrid models (also called hybrid effective field theory, or HEFT) use the tracer--matter connection as specified by LPT with displacements that are accurate at small scales from $N$-body simulations. Their combination leads to a powerful field-level description of the building blocks of structure formation. Recent applications of Hybrid EFT include: constructing emulators of clustering and lensing which are accurate to $k\simeq 0.6 \ihmpc$ \cite{Kokron_2021,hadzhiyska2021hefty,zennaro2021bacco}, modelling higher order statistics beyond the power spectrum \cite{banerjee2021modeling}, and characterizing the tracer--matter connection of simulated samples of galaxies \cite{zennaro2021priors,Kokron:2021faa}. \par
In this work we propose a novel way to leverage the intimate connection between Lagrangian perturbation theory and $N$-body simulations in order to improve the precision and accuracy of simulation-based predictions significantly. Specifically, we utilize the principle of control variates in order to create realizations of surrogate Universes with the same large-scale noise as those measured in $N$-body simulations. We use first order LPT (also known as the Zel'dovich Approximation \cite{1970A&A.....5...84Z}) to analytically predict the means of summary statistics. While the use of control variates in cosmology is not new \cite{chartier2020,chartier2021,Chartier:2022kjz}, previous work has relied on running ensembles of simulations in order to employ this technique. Our methods, in contrast, are computationally inexpensive. They rely only on data outputs that are a standard part of producing initial conditions for cosmological simulations, and an additional post-processing step that is identical in the full simulation and in the surrogate simulation. \par 
This paper is structured as follows: in \S~\ref{sec:controlvariates} we review the control variates technique, the variance reduction tool we have adopted in this publication. We discuss how this variance reduction is driven by the cross-correlation coefficient between the expensive simulation we wish to improve and the surrogate version we run. We also discuss how even the simplest rendition of LPT, the Zel'dovich approximation (ZA), produces Universes which are highly correlated with the results of full $N$-body simulations at low redshifts. In \S~\ref{sec:lpt} we give a brief overview of LPT and how it is used to predict the summary statistics of biased tracers. We focus on how to predict observables within LPT using both analytic calculations as well as grid-based realizations of the same expressions. In \S~\ref{sec:pijcv} we then proceed to re-formulate the control variates problem within the context of improving measurements of the basis spectra that make up the two-point statistics of biased tracers in LPT. We present the results of our implementation of control variates for paired ZA realizations in \S~\ref{sec:results}. We apply our methodology to three classes of tests, of increasing complexity. We use $N=100$ high-resolution $N$-body simulations, as well as paired ZA realizations, to assess the statistical performance of our technique. We begin by applying control variates to the matter power spectrum, extend to the statistics of a sample of galaxies populated by a halo occupation distribution (HOD) procedure, and conclude by looking at all ten basis spectra that span second order Lagrangian bias models.
Additionally, we quantify the reduction in variance from this technique from our ensemble of  simulations and consistently find strong reduction in uncertainty for all of the forms of power spectra assessed in this work. The improvements range from a factor of 10$\times$ reduction to nearly 1000$\times$ depending on the specific basis spectrum in question. We also discuss potential applications of these techniques beyond just improving predictions from cosmological emulation boxes. In \S~\ref{sec:conclusions} we summarize our results and identify promising future directions. \par 
Our appendices discuss technical aspects of this technique in order to ensure a simulation will benefit from its use. In Appendix~\ref{appendix:beyondloop} we derive the necessary expressions to model the power spectrum of biased tracers in the Zel'dovich approximation, which is crucial in ensuring the success of our techniques. In Appendix~\ref{appendix:calibrationsuite} we discuss the requirements imposed on simulations in order to ensure their paired ZA realizations will accurately match analytic predictions. In Appendix~\ref{appendix:beyondza} we discuss the potential of extending our techniques beyond the Zel'dovich approximation, as well as challenges that must be circumvented before this can be achieved in practice.

\section{Control variates and variance reduction}
\label{sec:controlvariates}

Control variates are a statistical technique employed to reduce the variance of quantities estimated with limited samples of data. They're applicable when one can create correlated approximate realizations with well-characterized means \cite{mcbook}. In the following sections we give a brief introduction to the theory of control variates and discuss their current applications within a cosmological context, as well as limitations to the technique as it is currently formulated. We will propose that the Zel'dovich approximation can be used as a control variate, and explore its correlation with the matter density field. In cosmology, control variates have recently been applied in the \texttt{CARPool} technique \cite{chartier2020,chartier2021,Chartier:2022kjz} as well as reducing the variance in statistics measured from the \texttt{AbacusSummit} \cite{Maksimova:2021ynf} suite of simulations \cite{Ding:2022ydj}. \par 

\subsection{Standard control variates}

Say we are interested in precisely estimating the mean of a simulated observable, $\hat{x}$.  Suppose also that we have a related quantity, the \emph{control variate} $\hat{c}$, that is significantly cheaper to produce than $\hat{x}$ but is correlated with it.  Then, we may define a quantity
\begin{equation}
    \hat{y} \equiv \hat{x} - \beta ( \hat{c} - \mu_c) \quad ,
\label{eqn:controlvariate}
\end{equation}
with $\mu_c = \langle \hat{c} \rangle$.  Taking the expectation value of Eq.~\ref{eqn:controlvariate} shows that $\langle\hat{y}\rangle$ is an unbiased estimator of $\langle \hat{x} \rangle$ for any $\beta$. The covariance between $\hat{c}$ and $\hat{x}$ can allow ${\rm Var}(\hat{y}) < {\rm Var}(\hat{x})$ and in fact ${\rm Var}(\hat{y})$ can be minimized by taking
\begin{equation}
    \beta^{\star} = \frac{{\rm Cov}[\hat{x},\hat{c}]}{{\rm Var}[\hat{c}]}.
\end{equation}
This value of $\beta^\star$ leads to a variance reduction 
\begin{equation}
    \frac{{\rm Var}[\hat{y}]}{{\rm Var}[\hat{x}]}
    = 1 - \frac{{\rm Cov}^2[\hat{x},\hat{c}]}{{\rm Var }[\hat{x}]{\rm Var}[\hat{c}]}
    = 1 - \rho_{xc}^2,
\label{eqn:variancered}
\end{equation}
where we've defined the cross-correlation coefficient $\rho_{xc} = {\rm Cov}[\hat{x}, \hat{c}]/(\sigma_x \sigma_c)$. Therefore, using a highly correlated surrogate that is inexpensive to produce can lead to significant improvements in estimation of such quantities without having to produce many realizations of $\hat{x}$, which could be computationally expensive. Within the context of cosmology, $\hat{x}$ can be the power spectrum of an $N$-body simulation \cite{chartier2020}, or a quantity such as the covariance matrix of that power spectrum \cite{chartier2021}. Our interest will be in the case when $\hat{x}$ is a given element of the basis spectrum $P_{ij} (k)$, which we define shortly in \S~\ref{sec:lpt} . The basis spectra $P_{ij}(k)$ are the building blocks of the statistics of biased tracers within LPT. This scenario includes as special cases the matter power spectrum and the spectrum of a biased tracer (e.g.\ galaxies).\par
One of the key limitations in applying control variates to problems in computational cosmology so far has been the computational cost of estimating the mean $\mu_c$ of the variate. If improperly estimated, the estimator $\hat{y}$ could become biased and the variance reduction becomes significantly hindered by the variance associated with estimating $\mu_c$. Even when using fast approximate $N$-body solvers such as \texttt{FastPM} \cite{Feng2016} or \texttt{COLA} \cite{Tassev_2013,tassev-scola}, the computational cost incurred from having to run many realizations can quickly limit possible gains from employing the technique. Another limitation comes in estimating the optimal form of $\beta^\star$, especially in the multivariate problem. This involves computing many realizations of $\hat{x}$ and $\hat{c}$ in order to estimate the full ${\rm Cov}[\hat{x},\hat{c}]$ matrix across multiple modes. When $\hat{x}$ is an observable measured from a cosmological $N$-body simulation, the number of realizations for a \emph{single} cosmology required to estimate a numerically stable $\beta^\star$ can easily exceed the total number of simulations typically produced for a whole emulation suite.

\begin{figure}
    \centering
    \includegraphics[width=\textwidth]{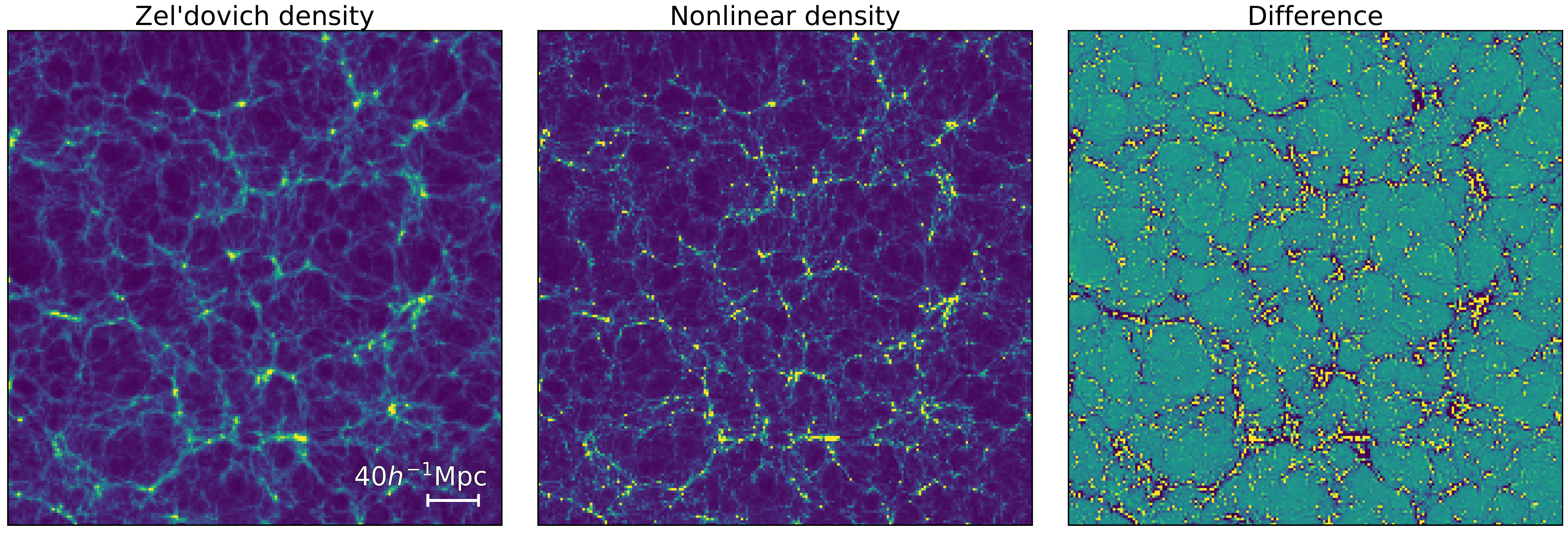}
    \caption{Comparison between the distribution of particles from a full $N$-body simulation and a matched realization using the Zel'dovich approximation, at $z=0.5$. The panels showing the Zel'dovich and nonlinear density fields $\delta^{ZA}$ and $\delta^N$, respectively, range from [-1,  $4\delta_c$] with $\delta_c = 1.686$ the critical threshold density for spherical collapse. The right-most panel shows the difference between the two fields, with the color bar for the rightmost plot encompassing the range $\delta^{ZA}(\bx) - \delta^{N}(\bx) \in [-1, 1]$. The maps shown are roughly $390 \times 390\, h^{-1}$Mpc, projected across roughly $20h^{-1}$Mpc.}  
    \label{fig:densplot}
\end{figure}
\subsection{The Zel'dovich approximation as a control variate}
It is highly desirable to define a control variate which simultaneously correlates significantly with the $N$-body statistic, but whose analytic properties are known exactly. As alluded to in the introduction, cosmological $N$-body simulations are often initialized using LPT. The first order solution in LPT is also known as the Zel'dovich approximation \cite{1970A&A.....5...84Z,White:2014gfa}. While we will describe LPT and the Zel'dovich approximation in further detail in \S~\ref{sec:lpt}, the approximation qualitatively states that fluid elements simply move in straight lines through the Universe with a displacement that is proportional to the linear growth factor. The direction of these displacements is seeded by the initial density fluctuations, and is found by solving for the linearized continuity equation in Fourier space. The properties of density fields in LPT are known to arbitrary precision, as they are analytic in nature. It is instructive, then, to consider whether the same theory used to initialize an $N$-body simulation can be extended to lower redshifts and used as a control variate. In Fig.~\ref{fig:densplot} we show 20 $h^{-1}$Mpc projections of the Zel'dovich and $N$-body density fields, which share the same initial conditions, evolved to $z=0.5$. As one can see, the Zel'dovich approximation captures the structure of the cosmic web in a striking fashion. The initial conditions and non-linear distributions used come from the \quijote \cite{Villaescusa_Navarro_2020} `high resolution' (HR) suite of dark matter-only $N$-body simulations, which have a volume of $V = 1 (h^{-1} {\rm Gpc})^3$ and $N = 1024^3$ particles. The fundamental grid size of these simulations is given by $L_{\rm cell} \approx 1 h^{-1}{\rm Mpc}$.  \par 
If the correlation between the Zel'dovich density and the non-linear density extends itself beyond the visual correlation of Fig.~\ref{fig:densplot}, a promising picture of its use as a control variate arises. The Zel'dovich approximation is a surrogate for structure formation that is typically available for all simulations and whose analytic results are known exactly. The basis spectra $P_{ij} (k)$ that are the building blocks for clustering and lensing statistics can be computed with no approximations within Zel'dovich. At the same time, Zel'dovich ``realizations'' of full $N$-body simulations are extremely inexpensive to generate (usually using codes that are already part of the N-body pipeline), and form a field-level description for \emph{any} summary statistic that one wishes to measure. Thus, if one can ensure that the two different approaches (analytic and grid-based) to calculating statistics in the Zel'dovich approximation are in numerical agreement, then the control variates technique can be employed with negligible computational overhead to greatly improve the precision of measurements of non-linear basis spectra. \par 

\begin{figure}
    \centering
    \includegraphics[width=\textwidth]{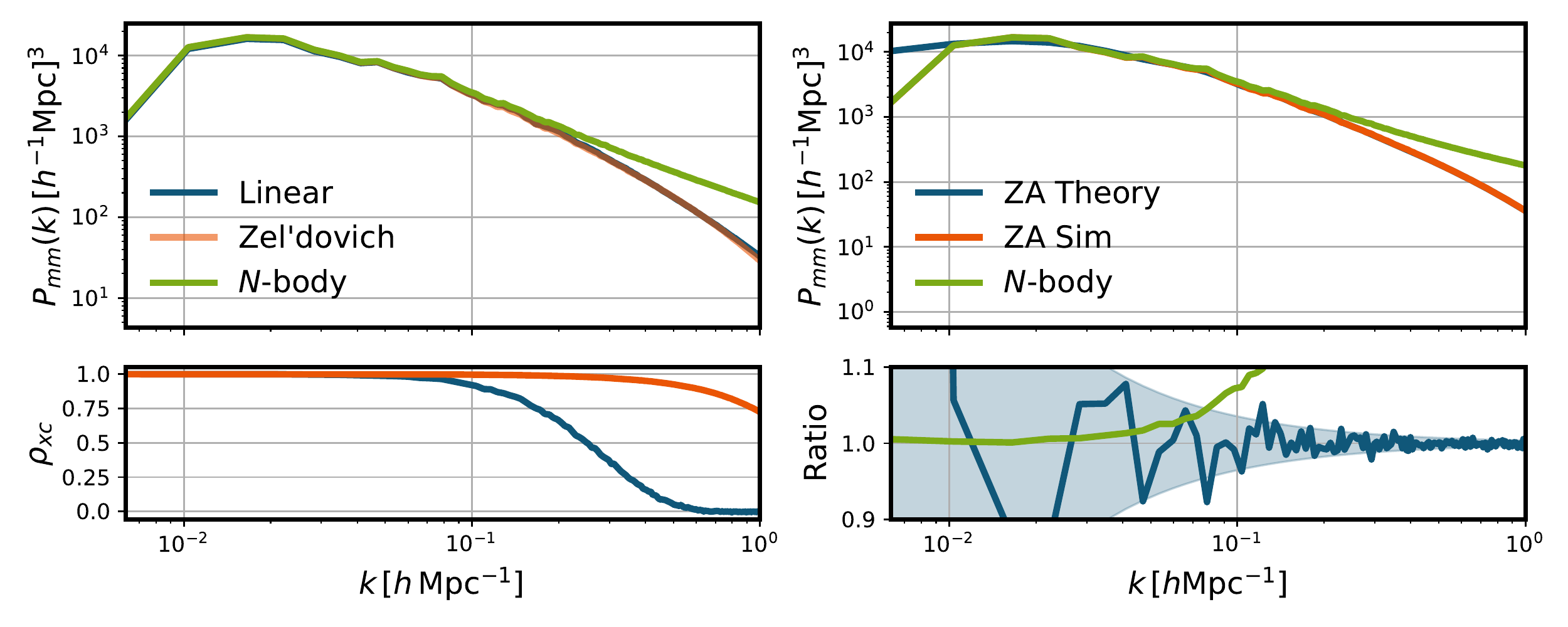}
    \caption{Comparison of the $z=0.5$ matter-matter power spectrum between linear theory, simulated ZA, analytic ZA and the full non-linear prediction from an $N$-body simulation. \emph{Top left:} Matter power spectra, $P_{mm}$, for the linear density field that seeded the initial conditions, the Zel'dovich-displaced density field, and the fully non-linear density field at $z=0.5$. \emph{Bottom left:} Cross-correlation coefficients between the non-linear density field and the linear (in blue) and Zel'dovich (in orange) densities, respectively. We see that despite having a similar power spectrum to the linear field, the Zel'dovich realization is much more highly correlated with the non-linear density down to small scales. The cross-correlation coefficient is $90\%$ at $k = 0.64\, h{\rm Mpc}^{-1}$ and $75\%$ at $k=1.0\, h{\rm Mpc}^{-1}$, while the linear density rapidly decorrelates. \emph{Top Right:} Matter power spectrum for the same Zel'dovich and non-linear densities, but showing in blue the result of the analytic Zel'dovich spectrum. \emph{Bottom Right:} Ratio between the two schemes used to make ZA predictions, as well as the ratio between the $N$-body and Zel'dovich realizations with the same initial conditions. The shaded region denotes 2-$\sigma$ error bars assuming a Gaussian disconnected covariance for the power spectrum. We see that the analytic and realization-based Zel'dovich calculations agree to high accuracy down to scales of $k_{\rm max} \approx 1 h {\rm Mpc}^{-1}$.
    }
    \label{fig:crosscorr}
\end{figure}
A more quantitative comparison elucidates the structure of this correlation. Specifically, we compute the cross-correlation coefficient between the density fields
\begin{equation}
    \rho_{xc} = \frac{P^{{\rm ZA},\, N}(k)}{\sqrt{P^{ N,N}(k) P^{\rm ZA,ZA}(k)}},
\end{equation}
as well as the power spectra of these fields, and show them in the bottom left panel of Fig.~\ref{fig:crosscorr}. We observe a strong correlation coefficient between the Zel'dovich and nonlinear densities, which is still at 75\% for $k=1 \ihmpc$. The linear density distribution, by comparison, reaches 75\% correlation with the nonlinear density at $k\approx 0.15 \ihmpc$.\par 
The fact that the Zel'dovich approximation produces distributions of matter that correlate highly with the non-linear matter distribution is not new. Indeed, this has been explored previously in many publications \cite{10.1093/mnras/260.4.765, Melott:1994ah,tassev2012,Tassev:2013rta, schmidt2021}. That the cross-correlation with the Zel'dovich approximation is high while the linear theory prediction is low, despite having similar two-point statistics, is due to the nature of the comparison made. In a Universe with only linear displacements, fluid elements will move across distances of roughly $\Sigma_\Psi$, where $\Sigma_\Psi^2 = \int dk P(k)/6\pi^2$. Comparing the fields before and after these displacements leads to decorrelation on scales smaller than $\Sigma_\Psi$, as dark matter particles have been displaced by this distance from their initial conditions. The chief impact of gravitational non-linearities is to introduce accelerations and slight deviations in the motions of fluid elements on scales smaller than this dispersion, and thus the Zel'dovich approximation describes most of this displacement. Curiously, it has also been noted that higher-order Lagrangian perturbation theories correlate \emph{more poorly} with the density field at lower redshifts, despite describing the matter power spectrum more accurately. \par
In this section we reviewed the method of control variates and its applications so far in cosmology. We pointed out the expense of estimating $\langle \hat{c} \rangle$ and $\beta$ when $\hat{c}$ is a fast $N$-body simulation as one of the main challenges of deploying this technique today. We proceeded to suggest that the Zel'dovich approximation, whose ingredients are generated when initializing any $N$-body simulation, could be a powerful control variate due to its high correlation with the non-linear density field.\par
We now proceed with a discussion of computing power spectra in LPT to \S~\ref{sec:lpt}. We leave Figures~\ref{fig:densplot} and ~\ref{fig:crosscorr} as tantalizing figures that point to the power of the Zel'dovich approximation as a control variate. We will formulate in \S~\ref{sec:pijcv} the control variate problem for the basis spectra of biased tracers, which will include as special cases both the power spectrum of matter density fluctuations and the statistics of \emph{any} biased tracer measured from simulations.

\section{Lagrangian Perturbation Theory Two Ways}
\label{sec:lpt}
In the Lagrangian picture of structure formation, the density contrast field is calculated from the movement of Lagrangian fluid elements across cosmic time. That is, particles located at a position $\bq$ at initial conditions are advected by a displacement $\bPsi$ to their final position
\begin{equation}
    \bx (a) = \bq + \bPsi (\bq, a).
\end{equation}
where $a=(1+z)^{-1}$ is the scale factor.
If the initial distribution of densities is approximately uniform, $\rho (\bq) \approx \bar{\rho}$, then at late times the mapping from Eulerian to Lagrangian coordinates gives an evolved density distribution from the continuity equation
\begin{equation}
    \label{eqn:lptdelta}
    1 + \delta (\bx, a) = \int d^3q\, \delta^D (\bx - \bq - \bPsi(\bq, a)),
\end{equation}
where $\delta^D$ is the Dirac delta function. In the presence of biased tracers, Eqn.~\ref{eqn:lptdelta} is extended by including a functional $F[\delta (\bq)]$ that specifies the tracer--matter connection at early times. The components of this functional are similarly advected, and the late-time tracer density $\delta_t (\bx, a)$ is given by
\begin{equation}
    \label{eqn:lpttracer}
        1 + \delta_t (\bx, a) = \int d^3q\, F[\delta (\bq)] \delta^D (\bx - \bq - \bPsi(\bq, a)).
\end{equation}
The functional $F[\delta(\bq)]$ is normally expanded to second order as \cite{Matsubara2008,Vlah_2016,Desjacques:2016bnm}
\begin{align}
\label{eqn:biasexp}
    F[\delta (\bq)]  \approx\, &1 + b_\delta \delta (\bq) + b_{\delta^2} (\delta^2(\bq) - \langle \delta^2 \rangle )\,+\\
\nonumber & b_{s} (s^2(\bq) - \langle s^2 \rangle)+ \,b_{\nabla^2}\nabla^2 \delta(\bq) + \cdots
\end{align}
where $s^2 (\bq) = s_{ij} (\bq) s^{ij} (\bq)$ is the tidal field strength, and $s_{ij}(\bq)$ is the tidal tensor defined as 
\begin{equation}
    s_{ij} (\bq) = \left ( \frac{\partial_i \partial_j}{\partial^2} - \frac{\delta_{ij}}{3} \right ) \delta(\bq).
\end{equation}
Advecting each component of Eqn.~\ref{eqn:biasexp} individually leads to the description of the late-time tracer field in terms of a set of advected operators 
\begin{align}
    \label{eqn:latetime}
    \delta_{t} (\bx) &= \delta_m(\bx) + b_{\delta} \mathcal{O}_\delta (\bx) + b_{\delta^2} \mathcal{O}_{\delta^2} (\bx) + b_{s} \mathcal{O}_{s^2} (\bx) + b_{\nabla^2} \mathcal{O}_{\nabla^2 \delta} (\bx) + \epsilon (\bx) \\
    &=  \sum_{ i \in \{m, \delta, \delta^2, \cdots\}} b_i \mathcal{O}_i + \epsilon(\bx),
\end{align}
where $\mathcal{O}_i$ are the advected operators and $\epsilon (\bx)$ is a stochastic field which quantifies both inherent randomness in the process of tracer formation as well as the impact of neglected higher-order operators. Note that in this notation $\mathcal{O}_m(\bx) = \delta_m(\bx)$ and $b_m\equiv 1$.  We further follow the convention in the LPT literature and use as aliases $b_1\equiv b_{\delta}$ and $b_2\equiv b_{\delta^2}$.
Tracer power spectra will receive contributions from correlations between the advected operators that compose the functional $F$. This decomposition will have the form
\begin{align}
\label{eqn:basisspec}
    &P^{tt} (k) = \sum_{i,j \in \{m, \delta, \delta^2, \cdots \}} b_i b_j P_{ij}(k), \\
    &P^{tm} (k) = \sum_{j\in \{m, \delta, \delta^2, \cdots \}} b_j P_{mj} (k),
\end{align}
where $P_{ij}$ is the cross-spectrum $\langle \mathcal{O}_i \mathcal{O}_j\rangle$ and $P_{mj}$ are cross-correlations between the matter density field and the bias operators\footnote{In the main text we neglect the effect of massive neutrinos, which affect the clustering of matter and galaxies in distinct ways --- while neutrinos contribute to the matter field, to a very good approximation galaxies trace the baryon-cold dark matter field $\delta_{cb}$ \cite{Castorina15}, which clusters on small scales (unlike the neutrinos, $\delta_\nu$). In order to accurately capture this effect we need to distinguish between two ``m'' fields: $\delta_{cb}(\bx)$, the overdensity of $cb$ particles obtained by advecting them via N-body displacements, and $\delta_m (\bx)$, the nonlinear matter field obtained by mass weighting the $cb$ and $\nu$ particles. All of the bias operators should be built from the initial $\delta_{cb}(\bq)$. Then we simply swap in these two versions of the ``m'' term in the above equations, e.g.\ for the cross spectrum
$P^{tm}(k)=\sum_j b_j P_{mj}(k)$ where $j\in \{cb, \delta_{cb}, \delta_{cb}^2, \cdots \}$.}.  Since the matter field is obtained by weighting each particle with weight 1, the matter power spectrum ($P_{mm})$ is often written as $P_{11}$ in the language of Lagrangian bias. For this work, we'll neglect the $\nabla^2 \delta (\bq)$ operator in our analysis and only work with $\{b_1, b_2, b_s\}$ as bias parameters. The reasons are two-fold; we expect that basis spectra of the type $P_{\nabla^2\delta X}$ will scale as $-k^2 P_{mX}$ analytically \cite{Kokron_2021}, and numerically realizing the $\nabla^2 \delta(\bq)$ fields is challenging due to the extreme sensitivity of the field to small scales (high $k$). More careful numerical studies of the $\nabla^2 \delta (\bq)$ field in the context of hybrid EFT models are of great interest and will be explored in future work.  \par
In Lagrangian Perturbation Theory, the displacements $\bPsi$ whose correlations serve as input for predictions of the spectra $P_{ij}$ are computed order-to-order in perturbation theory. To third order, this expansion is commonly written as \cite{Michaux:2020yis} 
\begin{align}
\label{eqn:psi3}
    \bPsi^{\rm 3LPT} (\bq, a) \approx D(a) \bPsi^{(1)}(\bq) + D^{(2)}(a) \bPsi^{(2)}(\bq) + D^{(3)} (a) \bPsi^{(3)}(\bq),
\end{align}
where $D^{(n)}(a)$ is the $n$-th order solution to the growth factor in perturbation theory \cite{Bernardeau:2001qr}. At any given order $(n)$, $\bPsi^{(n)}$ can be computed from suitable convolutions of $n$ powers of the first-order solution to displacements, also known as the Zel'dovich approximation. Today, efficient codes exist that implement the analytic equations for one-loop (combined fourth-order) power spectra in a numerically efficient manner, such as \texttt{velocileptors}\footnote{Available at \href{https://github.com/sfschen/velocileptors}{https://github.com/sfschen/velocileptors}.} \cite{Chen_2020}, and also codes that compute higher-order initial conditions for $N$-body simulations by numerically evaluating the displacements in Eqn.~\ref{eqn:psi3} such as \monofonic  \cite{Michaux:2020yis} and \texttt{LEFTfield} \cite{schmidt2021}\footnote{Available at \href{https://bitbucket.org/ohahn/monofonic}{https://bitbucket.org/ohahn/monofonic} and \href{https://gitlab.mpcdf.mpg.de/leftfield/release/leftfield}{https://gitlab.mpcdf.mpg.de/leftfield/release/leftfield}, respectively.}. For the rest of this text we will focus on only the Zel'dovich approximation, and defer a discussion of implementing our methodology for higher order Lagrangian Perturbation Theory in Appendix~\ref{appendix:beyondza}.\par 
\subsection{Analytic predictions in the Zel'dovich approximation} 
In the first order solution to Lagrangian Perturbation Theory, the Zel'dovich approximation, fluid elements in the Universe propagate in straight lines, with a direction set by the potential sourced by the initial matter distribution. These displacements are obtained by solving the linearized continuity equation, and read
\begin{equation}
\label{eqn:zapsi}
    \Psi^{\rm ZA}(\boldsymbol{q}, a) = D(a) \int \frac{d^3 k}{(2\pi)^3} e^{i \boldsymbol{k} \cdot \boldsymbol{q}} \frac{i \boldsymbol{k}}{k^2} \delta (\boldsymbol{k}).
\end{equation}
The late-time tracer density, in Fourier space, is then given by the Fourier transform of Eqn.~\ref{eqn:lptdelta}
\begin{align}
\label{eqn:deltak}
     \delta_t(\bk) &= \int d^3q\, e^{i \bk \cdot \bq} \left [ F[\delta (\bq)]e^{i \bk \cdot  \bPsi (\bq)} - 1 \right ] \\
     &\equiv \sum_{i\in \{m, \delta, \delta^2, \cdots \}} b_i \mathcal{O}_i (\bk),
\end{align}
where we've defined the \emph{advected} operators $\mathcal{O}_i (\bk)$
\begin{equation}
\label{eqn:oi}
    \mathcal{O}_i (\bk) \equiv \int d^3q \, e^{i \bk \cdot (\bq + \bPsi (q))}  F_i (\bq),
\end{equation}
and $F_i (\bq)$ is a Lagrangian element of the functional that composes Eqn.~\ref{eqn:biasexp}.
The right-most term of Eqn.~\ref{eqn:deltak} can be neglected, as it only contributes when $\bk = 0$. The tracer-tracer power spectrum may then be written as \cite{Taylor_1996,Matsubara2008,Carlson_2012}
\begin{align}
    P^{tt} (k) &= \sum_{i,j} b_i b_j \langle \mathcal{O}_i (\bk) \mathcal{O}^*_j (\bk) \rangle \\
    &= \sum_{i,j} b_i b_j \underbrace{\int d^3q e^{i \bk \cdot \bq} \left \langle F_i (\bq_1) F_j (\bq_2) e^{i \bk \cdot \Delta (\bq)} \right \rangle}_{P_{ij}(k)},
\end{align}
where we have defined $\Delta (\bq) = \bPsi (\bq_1) - \bPsi(\bq_2)$ and $\bq \equiv \bq_1 - \bq_2$\footnote{The integral for the basis spectra $P_{ij}(k)$ depend only on $\bq$ due to translation invariance.}.
The expectation value above is computed by using the cumulant expansion as well as defining appropriate source currents for the Lagrangian fields, $F_i(\bq)$. When working strictly within the Zel'dovich approximation, $\Delta (\bq)$ is a Gaussian random variable and only the second connected moment remains, significantly simplifying the calculation of the basis spectra $P_{ij}(k)$. \par 
In Appendix.~\ref{appendix:beyondloop} we provide a detailed derivation of the structure of each basis spectrum, $P_{ij}$, within the Zel'dovich approximation. We include several new terms that exist beyond the standard ``one-loop'' (quadratic in the two-point correlation function) order. We implement them in a new code, \zenbu\footnote{``Ze(ldovich calculations for) N-B(ody Em)u(lators)'', available at \href{https://github.com/sfschen/ZeNBu}{https://github.com/sfschen/ZeNBu}}. We use \zenbu\ for the analytic calculations in the rest of this paper unless otherwise specified. 

\subsection{Grid-based predictions in the Zel'dovich approximation}
\label{subsec:gridlpt}
An alternative way of computing the basis spectra $P_{ij}$ comes from employing \emph{grid-based} Lagrangian Perturbation Theory. Given a fixed realization of the initial Lagrangian density $\delta(\bq)$, one computes the displacements $\bPsi(\bq, a)$ by numerically evaluating the Fourier transform in Eqn.~\ref{eqn:zapsi}. Each element of the initial grid is then advected using these displacements, and they are used to reconstruct the final density field as in Eqn.~\ref{eqn:lptdelta}. Grid-based LPT has been well-studied in the context of setting up initial conditions for cosmological $N$-body simulations, with codes for ZA, 2LPT \cite{Crocce:2006ve} and 3LPT \cite{Michaux:2020yis} being widely available. Recent works have also investigated $n$-LPT at arbitrary order \cite{schmidt2021}, including to assess the fundamental convergence properties of LPT \cite{Rampf_2020}. Other applications of grid-based LPT include studying the fundamental limits of perturbative models \cite{Baldauf:2015zga} and using LPT as forward models to reconstruct the initial conditions of data taken from cosmic surveys \cite{Seljak_2017,Modi_2019,Modi_2021}. Relative to analytic approaches, grid-based schemes allow for ``field-level'' assessments of perturbation theories. Computing any summary statistic in grid-based PT is achieved by analyzing the output as a mock dataset. Grid-based schemes have also recently seen interest within the context of Eulerian perturbation theories \cite{Taruya_2018,Taruya:2020qoy,taruyafield}. However, Eulerian theory does not capture large-scale displacements which are naturally included in Lagrangian schemes. \par 

In the right panel of Fig.~\ref{fig:crosscorr} we show a comparison between the matter power spectrum measured from the Zel'dovich density field, the fully nonlinear field and the matter power spectrum as predicted by \zenbu. Both ZA predictions have the same Gaussian smoothing applied to their linear power spectra, at a scale of $k_{\rm cut} = \pi N_{\rm mesh} / L_{\rm box} \approx 3.2 \ihmpc$, given by Eqn.~\ref{eqn:cutoff}. The predictions agree exquisitely to small scales of $k = 1 \ihmpc$. \par 
Emulators based on Lagrangian Perturbation Theory (hybrid EFT) have recently been introduced as powerful models for describing the basis spectra of Eqn.~\ref{eqn:basisspec}, where the non-linear displacements from $N$-body simulations are used, as opposed to perturbative displacements \cite{modichenwhite19,Kokron_2021,zennaro2021bacco,hadzhiyska2021hefty}. While these emulators are promising tools, there are still limitations which have prevented their wide adoption. For example, most LPT-based emulators eventually revert to analytic predictions at large scales. This happens when sample variance in the suite of $N$-body simulations used to construct the emulator becomes too large, preventing accurate predictions from being made. Many algorithms for performing $N$-body simulations exhibit slight discrepancies in how they evolve large-scale growth \cite{Heitmann:2008eq,Zennaro_2016,Schneider_2016,Garrison:2016vvp,DeRose:2018xdj} relative to linear theory. A small-mismatch between LPT predictions and measurements in simulations can introduce unphysical features in emulated spectra\footnote{For example, refs.~\cite{Kokron_2021,hadzhiyska2021hefty} had to take care to connect LPT predictions to HEFT spectra without introducing discontinuities in the emulator predictions.}. These features, while local in Fourier space, significantly affect configuration space-based emulators using these same suites. Another challenge is the trade-off between covering a cosmological parameter space adequately while also having sufficiently large volumes that emulator accuracy is below the requirements for the next generation of large-scale surveys. Indeed, attempts to use \texttt{anzu}\footnote{Available at \href{https://github.com/kokron/anzu}{https://github.com/kokron/anzu}}, the emulator of ref.~\cite{Kokron_2021} based on the \aemulus suite \cite{DeRose:2018xdj}, in an analysis of data \cite{white2022cosmological} ran into the issue that the bounds of the emulator were too restrictive relative to the posteriors obtained from the analysis. \par
\begin{figure}
    \centering
    \includegraphics[width=\textwidth]{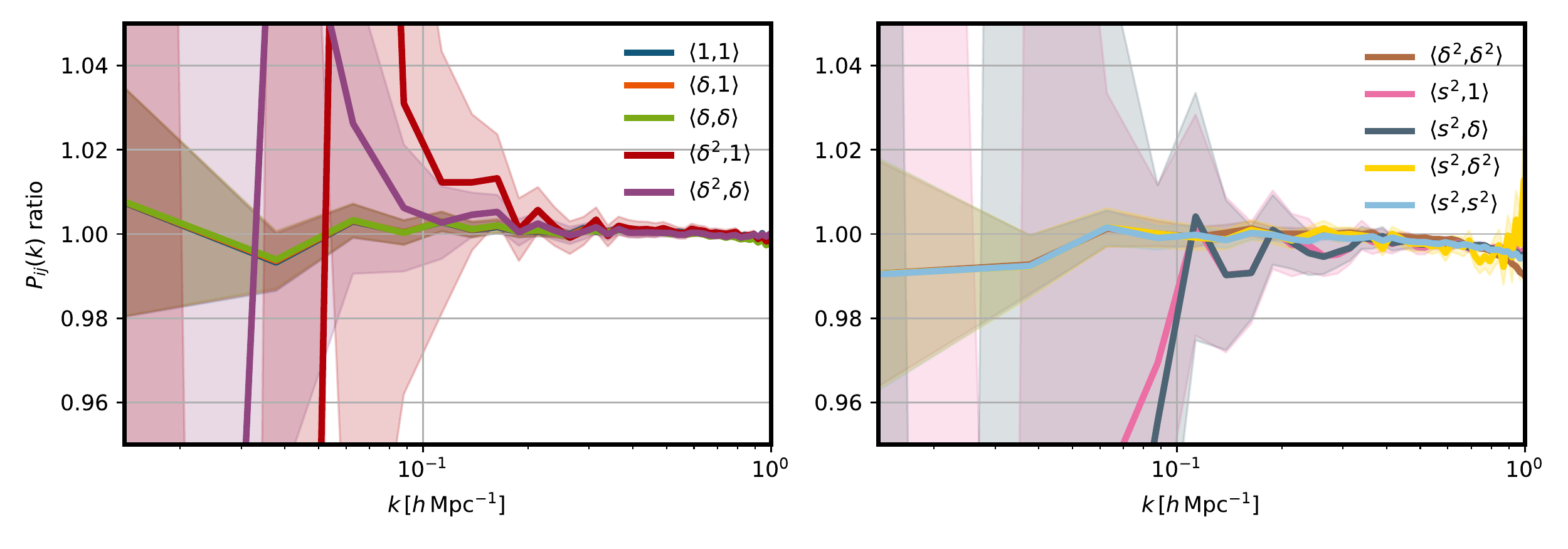}
    \caption{Ratio between the grid-based ZA basis spectra as measured from 100 \quijote HR boxes and spectra computed analytically in the Zel'dovich approximation. Each line corresponds to the cross-spectrum of the advected particles weighted by fields $(i,j)$ in the initial conditions. The measurements are binned with $\Delta k = 4$ modes, corresponding to approximately $\Delta k = 0.025 \ihmpc$. The shaded bands are the standard deviations measured from the Quijote HR boxes divided by $\sqrt{ 100}$.  }
    \label{fig:zaratio_bin}
\end{figure}

As such, reducing sample variance at large scales, especially in the transition regime between LPT and full non-linear displacements, for basis spectra measured from $N$-body based emulators could have the potential to enable larger parameter space coverage and more stable emulators than have been currently constructed. The striking correlation between the matter density fields constructed in the Zel'dovich approximation and the final result of the $N$-body simulation suggest that cheap Zel'dovich realizations can enable substantial reduction in sample variance of emulated quantities. This was pointed out, without control variates, in ref.~\cite{tassev2012} for the case of the matter power spectrum, but this principle holds for all basis spectra, as well as any other statistic that can be computed within the Zel'dovich approximation. We will illustrate how this is the case in the remainder of the text.
In Fig.~\ref{fig:zaratio_bin} we show a similar plot to the lower right-hand plot of Fig.~\ref{fig:crosscorr} but for all component spectra in ZA. We find near sub-percent agreement for almost all basis spectra to scales of $k=1 \ihmpc$. We note that the discrepancy for cubic basis spectra at large scales is due to these quantities being noisily measured despite using N=100 \quijote HR simulations. The small scale discrepancies for $\langle s^2 \delta^2 \rangle$ are due to issues in the smallest scales probed by our simulations\footnote{We discuss in Appendix~\ref{appendix:calibrationsuite} what is the dynamic range in scales that an $N$-body simulation should contain in order to accurately reproduce the analytic calculations of ZA.}. These small discrepancies are not an issue, as
\begin{enumerate}
    \item Contributions from higher order basis spectra are sub-leading relative to the full predictions of galaxy clustering and lensing. 
    \item Higher order basis spectra fall off rapidly at high $k$, and thus their difference will induce a very small bias in the variance-reduced estimate of that basis spectra.
    \item The regression coefficient $\hat{\beta}(k)$ will be small at high $k$, when the correlation between Zel'dovich and $N$-body is small. 
\end{enumerate}
The sub-percent agreement for most basis spectra to $k=1\ihmpc$ shown in Fig.~\ref{fig:zaratio_bin} shows there are no potential biases in using analytic calculations as the mean when applying ZA as a control variate for Lagrangian basis spectra. The use of analytic predictions for the mean of the control variate significantly reduces computational expenses in applying the technique. The other main potential challenge, as discussed, is computing the regression matrix $\beta$. We will discuss this in the following section.
\section{Control variates for LPT basis spectra}
\label{sec:pijcv}

Returning to the question of variance reduction, we now formulate the problem of control variates within the notation of basis spectra in LPT. The control variate problem as defined in Eqn.~\ref{eqn:controlvariate} may be written as\footnote{In principle one should consider the full multivariate control variate problem for the data vector $\{P_{ij}(k_1) \cdots P_{ij}(k_n)\}$, however for power spectrum estimation the diagonal approximation works reasonably well \cite{chartier2020}. }

\begin{equation}
\label{eqn:pkcv}
    \hat{P}_{ij}(k) = \hat{P}^{N}_{ij}(k) - \beta^{\star}_{ij} (k) \left (\hat{P}^{\rm ZA}_{ij} (k) - P^{\rm ZA}_{ij} (k) \right ),
\end{equation}
where $\hat{P}_{ij}^{ N}$ denotes the measured basis spectrum from full $N$-body simulations, $\hat{P}^{\rm ZA}_{ij}$ is the basis spectrum from ZA given the same initial conditions, and $P^{\rm ZA}_{ij}$ is the average ZA prediction, which can be computed either analytically or from several ZA realizations averaged together. The $\beta^{\star}$ term in this case is slightly more complicated. Assuming only contributions diagonal in Fourier wavenumber $k$ contribute we have 

\begin{align}
    \label{eqn:betacov}&{\rm Cov}[\hat{P}^{ N}_{ij} , \hat{P}^{\rm ZA}_{ij}](k) \propto \left \langle \hat{\mathcal{O}}_i^N (\bk) \hat{\mathcal{O}}_j^N (-\bk) \hat{\mathcal{O}}_i^{\rm ZA} (\bk) \hat{\mathcal{O}}_j^{\rm ZA} (-\bk) \right \rangle - P^N_{ij} (k) P^{\rm ZA}_{ij} (k),   \\
    \label{eqn:betavar}&{\rm Var}[\hat{P}^{\rm ZA}_{ij} ](k) \propto \left \langle \hat{\mathcal{O}}^{\rm ZA}_i (\bk) \hat{\mathcal{O}}^{\rm ZA}_j (-\bk) \hat{\mathcal{O}}^{\rm ZA}_i (\bk) \hat{\mathcal{O}}^{\rm ZA}_j (-\bk) \right \rangle - (P_{ij}^{\rm ZA}(k))^2. 
\end{align}
The operator $\hat{\mathcal{O}}_i(\bk)$ corresponds to the advected, late-time Lagrangian bias field whose cross-correlations form the basis spectra, as defined in Eqn.~\ref{eqn:oi}. When the ratio defining $\beta^{\star}$ is taken, we can neglect the proportionality constants related to the volume of the simulation box\footnote{There are also contributions which depend on the connected tri-spectrum between $N$ body and ZA fields which don't depend on the volume, but they are neglected when taking only disconnected terms.}. Since any choice of $\beta$ leads to an unbiased estimator, we adopt a compromise between the optimal $\beta^{\star}$ and a surrogate that is easier to compute while still providing substantial variance reduction. To do this, we will employ approximations in computing the covariances of Eqns.~\ref{eqn:betacov} and \ref{eqn:betavar}. Notably, we opt to select only the disconnected contributions from the above four-point correlations, which leads to a more amenable form
\begin{equation}
\label{eqn:betahat}
    \hat{\beta}_{ij}(k) \approx \frac{ \hat{P}_{ii}^{NZ} \hat{P}_{jj}^{NZ} + \hat{P}_{ij}^{NZ} \hat{P}_{ij}^{ZN}}{(\hat{P}_{ij}^{ZZ})^2 + \hat{P}_{ii}^{ZZ} \hat{P}_{jj}^{ZZ}},
\end{equation}
where $P_{ij}^{ZN}$ is defined as the spectrum $\langle \mathcal{O}_i^{\rm ZA} \mathcal{O}_j^N \rangle$. 

The quantity $\hat{\beta}$ is simple to compute from the simulation outputs used to generate the basis spectra of Lagrangian bias emulators. The main over-head is associated with producing the Zel'dovich-advected component fields. Additionally, since the estimator in Eqn.~\ref{eqn:betahat} is a ratio of power spectra measured from boxes which share the same initial phases, we expect $\hat{\beta}_{ij}$ to be a relatively noiseless quantity, despite depending on only a single ZA realization. We have verified that, while in principle the quantity $\hat{\beta}_{ij}(k)$ could vary from simulation to simulation the variance cancellation from the ratios of simulated quantities results in a measured $\hat{\beta}_{ij}$ with negligible variance. Additionally we note that, formally, using a $\hat{\beta}_{ij}(k)$ measured in the same simulation as the one whose variance is being reduced could affect the unbiased properties of the estimator in Eqn.~\ref{eqn:pkcv}. To avoid this issue, we use $\hat{\beta}_{ij}$ measured from Box 0 of \quijote HR as the regression matrix for all other boxes. \par
\begin{figure}
    \centering
    \includegraphics[width=\textwidth]{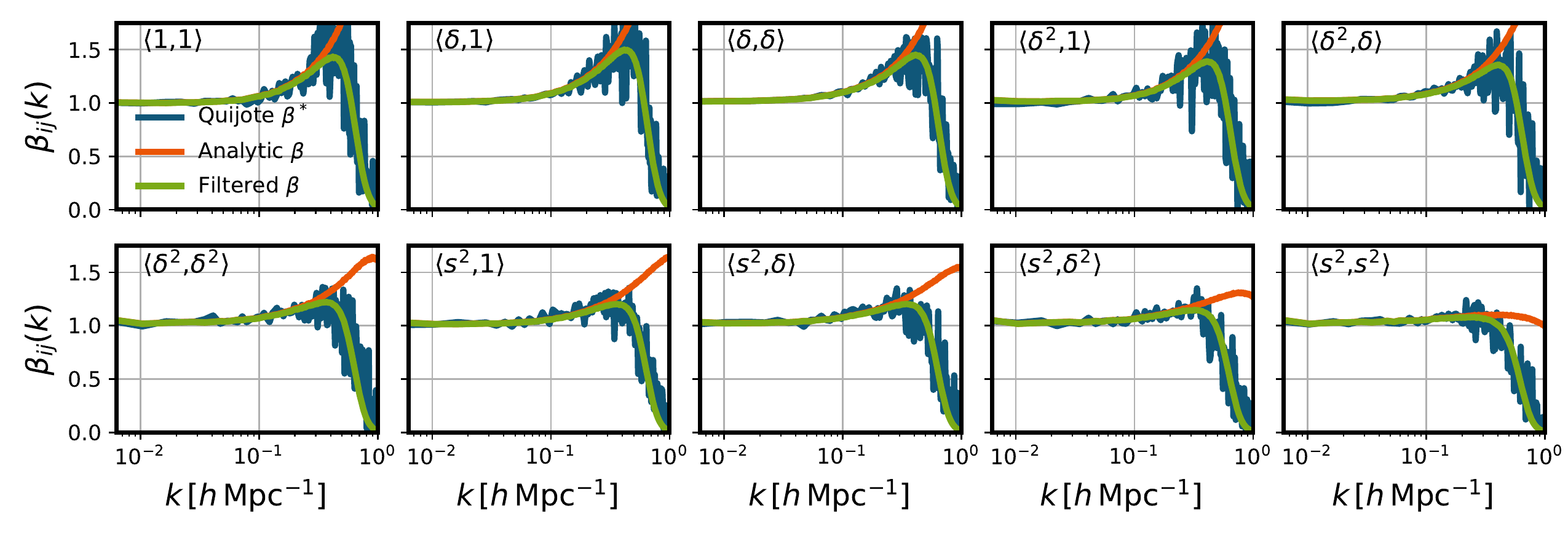}
    \caption{Plotting $\beta$ as measured empirically from \quijote HR, compared to the analytic approximation of Eqn.~\ref{eqn:betahat} as well as using the tanh filter function of Eqn.~\ref{eqn:tanhfilter} where $k_0$ and $\Delta_k$ have been calibrated to $\beta_{11}$. Despite calibrating to $\beta_{11}$ we find the filtered analytic $\beta$ are in good agreement with the empirical results for all other basis spectra.}
    \label{fig:cbetas}
\end{figure}
In order to test the approximation of $\hat{\beta}$ in Eqn.~\ref{eqn:betahat}, we also compute the numerical $\beta^\star$ from the \quijote HR boxes. We compute the covariance between ZA and $N$-body simulations explicitly and compute the multivariate estimator for $\beta^\star$ described in \cite{chartier2020}. In the context of basis spectra estimation, this $\beta^\star$ reads
\begin{align}
\label{eqn:betaempirical}
    \beta_{ij}^\star (k, k') = \sum_{k''} {\rm Cov}[\hat{P}^{ N}_{ij}(k) , \hat{P}^{\rm \rm ZA}_{ij} (k'')] {\rm Cov}[P_{ij}^{\rm ZA}(k''),P_{ij}^{\rm ZA}(k')]^{-1}.
\end{align}
The inverse covariance is computed using the Moore-Penrose pseudoinverse due to the small number of realizations. While the entire matrix $\beta^\star$ is poorly estimated with only $N=100$ simulations, we find that the diagonal component is relatively well measured. Thus, we proceed to compare our analytic approximation with the empirical result. This comparison is shown between the orange and blue lines of Fig.~\ref{fig:cbetas} for the ten basis spectra which span second order Lagrangian bias. 
The analytic approximation of Eqn.~\ref{eqn:betahat} holds well until scales of $k\approx 0.3 \ihmpc$, after which the empirically measured $\beta^\star$ damp to zero. This damping is not observed in the analytic approximation. The disagreement between the analytic form and our measurement can be understood by re-casting $\beta^\star$ as 
\begin{align}
    \beta^\star = \rho_{N, {\rm ZA}} \ \frac{\sigma_N}{\sigma_{\rm ZA}}.
\end{align}
The dominant contribution to the second term will scale as the ``transfer function'' between linear and non-linear power, and typically grows as a function of scale in $\Lambda$CDM. On the other hand, the disconnected approximation for $\rho_{N, ZA}$ we employ does not capture the full de-correlation between $N$-body dynamics and the Zel'dovich approximation. However, for all basis spectra we also find that this damping is very well approximated by a tanh function of the form
\begin{align}
\label{eqn:tanhfilter}
    F(k;k_0,\Delta_k) = \frac{1}{2} \left [ 1 - \tanh \left ( \frac{k - k_0}{\Delta_k} \right )   \right ].
\end{align}
Fitting values of $k_0$ and $\Delta_k$ to $\beta_{11}$, measured for the dark matter densities, we find fiducial values $k_0 = 0.618 \ihmpc$ and $\Delta_k = 0.167 \ihmpc$. These values give a very good description of high-$k$ damping for all other $\beta$ functions and we use them throughout the rest of this work. Thus, we adopt as an inexpensive approximation to the full empirical $\beta$ the product between Eqn.~\ref{eqn:betahat} and the filter function $F(k;k_0,\Delta_k)$. Choosing different values of $k_0$ and $\Delta_k$ also allows for control over the trade-off between any bias between the predicted and measured variate, and cancellation of sample variance. This damping of $\hat{\beta}$ also means any small-scale biases in our estimation of the mean Zel'dovich of the control variate can be safely neglected. In Fig.~\ref{fig:cbetas} we show in green the resulting damped $\beta$ for the ten basis spectra of Lagrangian bias. Despite calibrating the filter $F$ to the regression matrix of the matter power spectrum, its combination with the analytic approximation of Eqn.~\ref{eqn:betahat} provides a strong agreement for the diagonals of regression matrices of all 10 basis spectra. \par 
We have thus formulated the control variate problem for basis spectra in Lagrangian bias theory (which inclues the matter power spectrum and clustering/lensing of biased tracers as special cases) under the assumption the Zel'dovich approximation is a suitable control variate. We have derived an approximation for the regression matrix $\beta_{ij}$ which reproduces empirical results obtained from a suite of high resolution simulations. We therefore have all of the ingredients required to estimate $\hat{P}_{ij}(k)$ and quantify the variance reduction obtained from Zel'dovich control variates. 

\section{Results and discussion}
\label{sec:results}
In the following section we apply Zel'dovich control variates to progressively more complexcosmological power spectra. We begin with the matter power spectrum as a warm up, proceed to the case of extending the results for an HOD-like sample of galaxies, and conclude with the fully general case of improving measurements of all 10 basis spectra in second order Lagrangian bias models. \par 
All measurements in this section are carried out across the whole ensemble of $N=100$ \quijote HR simulations. While small in number relative to the full \quijote suite, we find that we need the higher dynamic range in order to match analytic and grid-based LPT, and prevent biases in the mean control variate estimate. We show why this is the case explicitly in Appendix~\ref{appendix:calibrationsuite}. We adopt as a fiducial choice the $z=0.5$ snapshot but note that there are no impediments to using other snapshots beyond computational costs. The use of $N=100$ simulations should allow, at least, for quantification of the reduction in the variance of numerical observables. We leave a quantification of their reduction of \emph{co-}variance to future work, as this is a significantly more numerically challenging problem.
\subsection{Variance reduction for the matter power spectrum}
\begin{figure}
    \centering
    \includegraphics[width=\textwidth]{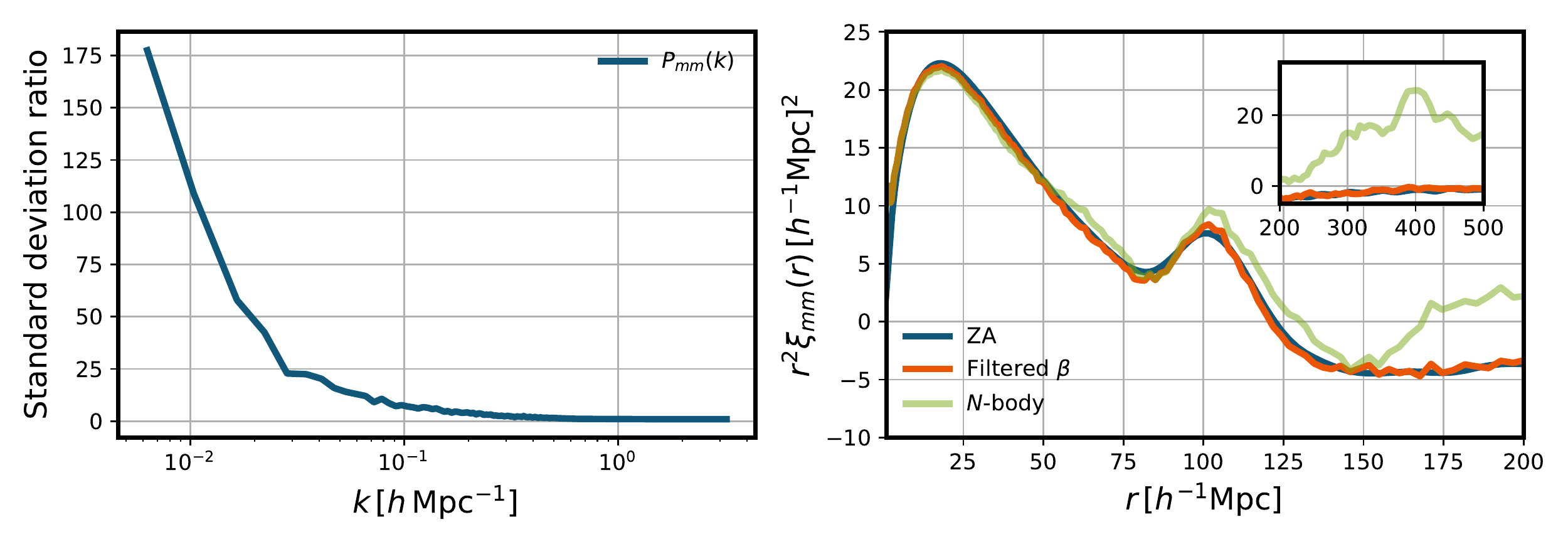}
    \caption{\emph{Left panel:} Scale-dependent ratio, between standard deviations of matter power spectra, from paired ZA control-variates and the empirical standard deviations from the N=100 \quijote HR boxes. \emph{Right panel:} Visualizing variance reduction in the configuration-space correlation function, $\xi_{mm} (r)$. The blue curve results from Hankel-transforming the Zel'dovich $P_{mm}(k)$ prediction from \zenbu. The orange and green curves are obtained by similarly Hankel-transforming empirically measured power spectra from a single \quijote HR box. The inset shows the increase in bias and variance for the standard $N$-body result at very large separations, while the control variate closely tracks the correct large-scale behavior predicted by the Zel'dovich approximation. 
    }
    \label{fig:pmm_reduce}
\end{figure}
In Eqn.~\ref{eqn:variancered} we derived that, in the standard control variates approach, the variance of the estimator is parametrically reduced by the cross-correlation coefficient between the full $N$-body result and the Zel'dovich surrogate. In Fig.~\ref{fig:crosscorr} we showed that for the case of the matter density field, this cross-correlation coefficient was substantial at $k=1\ihmpc$. However, Eqn.~\ref{eqn:variancered} assumes a form $\beta^\star$ which is different from the final form of $F(k;k_0,\Delta_k) \hat{\beta}_{ij} (k)$ that we used in this work. Thus, we turn to our statistical ensemble of simulations to quantify, in practice, how much variance reduction we can achieve through our technique.\par 
We begin by considering solely the matter density field and its auto-spectrum, $P_{mm}(k)$. In the language of Lagrangian bias this is the $P_{11}(k)$ spectrum, and we will show results for the whole suite of basis spectra shortly. The control variate estimator we've derived, explicitly, is given by

\begin{equation}
    \hat{P}_{mm} (k) = \hat{P}^N_{mm} (k) - F(k;k_0, \Delta_k)\left ( \frac{\hat{P}_{mm}^{ \rm * NZ}(k)}{\hat{P}_{mm}^{\rm * ZA}(k)} \right )^2 \left ( \hat{P}_{mm}^{\rm ZA}(k) - P_{mm}^{\rm ZA}(k) \right ),
\end{equation}
where $\hat{P}^{\rm *NZ}$ and $\hat{P}^{\rm *ZA}$ are spectra measured from a different box in order to prevent biasing the estimator, as discussed in \S~\ref{sec:pijcv}.
We measure the matter power spectrum in our Zel'dovich mocks, as well as in the full $N$-body fields, for all 100 boxes of \quijote HR. We also measure the cross-power spectrum between matter fields in all 100 matched boxes. To quantify the amount of variance reduction (or, equivalently, effective volume increase) we report the ratio of standard deviations
\begin{equation}
    \frac{\sigma^{\rm N_{body}}(k)}{\sigma^{\rm CV}(k)},
\end{equation}
as a function of scale obtained from implementing ZA control-variates for all basis spectra. These standard deviations are measured empirically from the \quijote HR suite. We show the result in the right-hand panel of Fig.~\ref{fig:pmm_reduce}. 

In agreement with the intuition of Eqn.~\ref{eqn:variancered}, we see that variance reduction is most substantial at large scales where the Zel'dovich approximation more faithfully captures the dynamics of structure formation. We also find that, due to the damping function $F(k;k_0, \Delta_k)$ at very small scales we observe no reduction in uncertainty from our estimator. At $k\simeq 0.01 \ihmpc$, the reduction in standard deviation is of the order of $\sigma^{\rm N_{body}}/\sigma^{\rm CV} \sim 110$. Since the standard deviation scales as
\begin{equation*} 
\sigma \propto V^{1/2},
\end{equation*}
this is equivalent to an effective volume increase of a factor of $\sim 10^4$. This reduction is larger at larger scales which are worse affected by sample variance. While the empirical $N$-body uncertainty grows as $k\to 0$, we find that for the paired ZA control variate the uncertainty scales as $\sigma^{\rm CV} \propto P(k)$, after $k\lesssim 0.03 \ihmpc$. This can be understood by treating the cross-spectrum between ZA and $N$-body as the same at large scales as the cross-spectrum between ZA and linear theory. Then, following ref.~\cite{Pontzen_2016} we can write $\rho_{\rm ZN}^2 = e^{- 2 (k/k_{\rm NL})^2}$. The end result is that $\sigma^2_{\rm CV} \propto P(k)^2 / (V \Delta k k_{\rm NL}^2)$ at large scales.\par 
To illustrate the degree of sample variance reduction we achieve, we also reconstruct the configuration-space correlation function $\xi_{mm} (r)$ directly from the empirically measured power spectra in boxes. We reconstruct the correlation function by evaluating
\begin{equation*}
    \hat{\xi}_{mm} (r) = \frac{1}{2\pi^2} \int_0^\infty k^2 dk\, j_0(kr) \hat{P}_{mm}(k),
\end{equation*}
which is done using \texttt{MCFit}\footnote{\href{https://github.com/eelregit/mcfit}{https://github.com/eelregit/mcfit}}. The empirically measured power spectra are extrapolated beyond $k \in \left [ k_f, k_{\rm Nyq} \right ]$\footnote{For \quijote HR, which has $L_{\rm box} = 1000 h^{-1}{\rm Mpc}$ and $N_{\rm mesh} = 1024$ we have $k_f \approx 6.3 \times 10^{-3} \ihmpc$ and $k_{\rm Nyq} \approx 3.2 \ihmpc$.} by constructing a linear spline in $(\log k, \log P(k))$, which is then re-binned in log-spaced $k$ bins. We show our reconstructed correlation functions in the left panel of Fig.~\ref{fig:pmm_reduce}, as well as the prediction from the Zel'dovich approximation. We find the reconstructed $\hat{\xi}$ from ZA control variates to have significantly reduced noise, as well as appropriate damping of the BAO. We also see that at very large scales $r \geq 10^2 h^{-1}{\rm Mpc}$ the result is in close agreement with linear theory, despite only extrapolating the power spectrum to lower $k$. In contrast, the result from attempting this $\xi_{mm}$ reconstruction for the default $N$-body result leads to significant biases beyond $r > 10^2 h^{-1} {\rm Mpc}$, as well as large variances around the trend line. We note that if we extend $\hat{P}(k)$ at $k < 2\pi/L_{\rm box}$ by including the noiseless $P(k)$ from linear theory, the reconstructed control variate $\xi$ does not change, while the $N$-body result becomes less biased at large scales but still with equivalently large variance. \par 
The reconstruction of large-scale $\hat{\xi}_{mm}(r)$ in this form, while a simple application, showcases the strength of this technique and the myriad ways in which the variance-reduced spectra could be used. We stress that the only additional steps in going from the green to the orange curves in Fig.~\ref{fig:pmm_reduce} are: running an inexpensive Zel'dovich mock at $z=0.5$ and from this measuring $P_{mm}^{\rm ZA}(k)$ and the cross-spectrum $P_{mm}^{\rm NZ}(k)$.
\subsection{Reducing the variance of biased tracer samples}
\begin{figure}
    \centering
    \includegraphics[width=\textwidth]{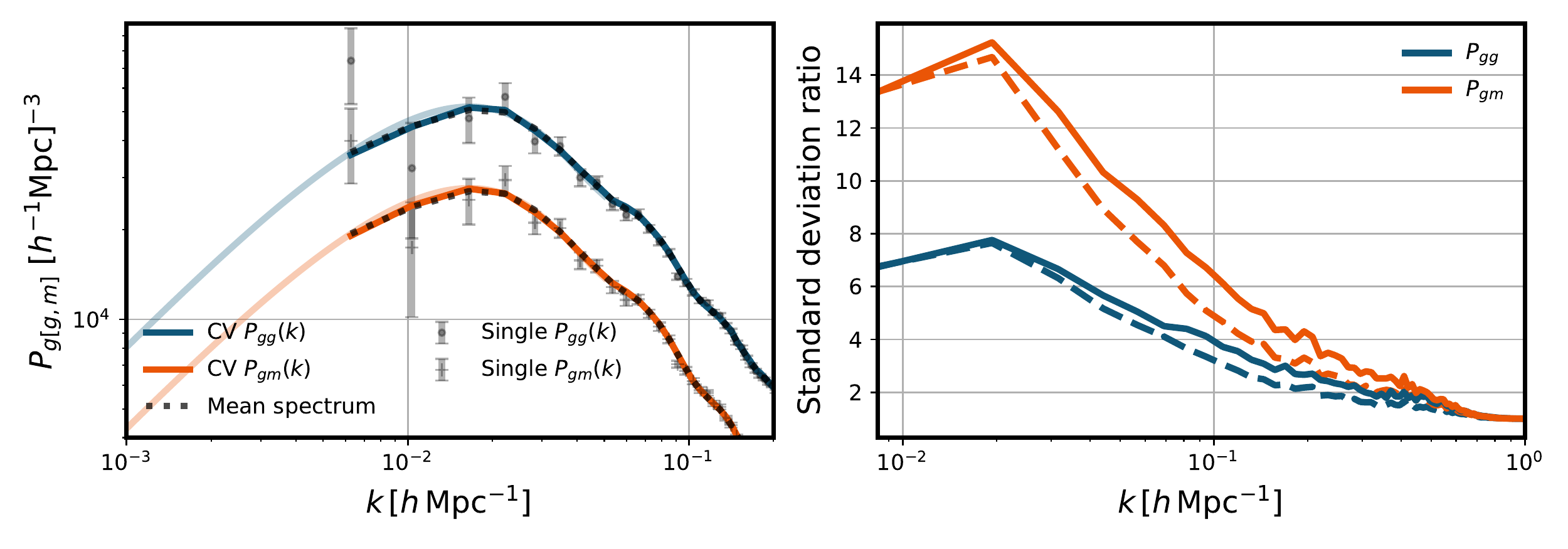}
    \caption{\emph{Left panel}: Measurements of tracer statistics with and without applying Zel'dovich control variates. The points show the power spectrum of galaxies populating halos in a randomly chosen \quijote HR box, with uncertainties estimated from the ensemble. The dark, solid colored lines show measurements from the same realization, but measured using control variates. Dotted black lines correspond to the mean of the \quijote HR ensemble of simulations. The transparent lines correspond to the Zel'dovich predictions, at large scales, using only $b_1$.  \emph{Right panel}: Ratio of the standard deviations of the spectra shown in the left panel. The dashed lines show the improvements in error reduction from only using matter field as a control variate, while the solid lines show the same technique but where we have also included the $b_1$ field in building the control variate.
    }
    \label{fig:tracer_reduce}
\end{figure}
Hydrodynamic simulations \cite{Dubois_2012, Le_Brun_2014,Schaye_2014,Khandai_2015,bahamas2017,Dav__2019,nelson2021illustristng} that try to reproduce galaxy formation ab initio, and radiative hydrodynamics simulations that try to understand the structure and evolution of reionization fronts \cite{Trac_2015} have extreme resolution requirements which significantly limit their volume. Measuring summary statistics in these simulations with less noise and to larger scales is thus highly desirable. Given the impacts of small-scale baryonic physics do not backreact on large scales, we can use control variates to extend their measurements to larger scales. Extending simulation-based models to larger scales is also highly desirable in light of recent advances in simulation--based inference. While the power spectrum at small scales possesses many modes, which require simulations to accurately describe, summary statistics at large scales provide complementary information and their combination can break parameter degeneracies that arise in complicated models. Purely simulation--based analyses of galaxy statistics such as those in \cite{Lange_2019,Yuan_2021,Lange_2021,Yuan:2022jqf,zhai2022} achieve impressive constraints on parameters, however are limited to fairly small scales ($r \lesssim 30 h^{-1} {\rm Mpc}$).\par 
We proceed to apply Zel'dovich control variates to a sample of ``galaxies'' selected from the \quijote HR boxes to demonstrate how our technique can extend the range of scales of measurements of biased tracer statistics in simulations. Using the public halo catalogs, we create density fields of halos, where each halo is weighted by the expected number of galaxies $\langle N(M) \rangle$ for its mass. The weights are derived from the halo occupation distribution of ref.~\cite{Zhai_2017}, which was fit to a sample of luminous red galaxies from the Sloan Digital Sky Survey.  Since control variates will be most powerful at large scales, this crude approximation to the galaxy--halo connection should be sufficient for the purposes of illustrating applications to tracer samples. While we have used catalogs of mock LRGs, these techniques are applicable to any tracer catalog from a simulation. This includes galaxies in hydrodynamical simulations, or the power spectrum of the electron distribution measured during reionization, for example.\par 
We wish to estimate the power spectra relevant to galaxy clustering and lensing, that is,
\begin{equation}
    P_{gg} = \langle \delta_g \delta_g\rangle,\quad P_{gm} = \langle \delta_g \delta_m \rangle
\end{equation}
in our simulations (having discussed $P_{mm}$ already). We first use the matter power spectrum itself as a surrogate as this is extremely cheap and quick to implement. We then define the control variate estimator for these spectra
\begin{equation}
    \hat{P}_{g[g,m]} (k) = P_{g[g,m]} - \hat{\beta}_{mm}^{*} \left [ \hat{P}^{\rm ZA}_{mm} (k) - P^{\rm ZA}_{mm} (k) \right ],
\end{equation}
where $P_{g[g,m]}$ is a short-hand for either spectrum used. While we expect that the cross-correlation coefficient between the matter power spectrum in Zel'dovich and the statistics of biased tracers to be lower than if we had used a surrogate for the tracer itself, we should still expect some degree of variance reduction. We show the results of this procedure in the right-hand panel of Fig.~\ref{fig:tracer_reduce}, in the dashed lines. Despite using a crude control variate, we still find substantial reduction in sample variance across all scales $k \leq 0.2 \ihmpc$. At $k=0.2 \ihmpc$ the autospectrum has a 2$\times$ reduction in its uncertainty, while for $P_{gm}$ we find a 3$\times$ reduction at this scale. The reduction is always higher for the galaxy--galaxy lensing spectrum $P_{gm}$, which can be understood from the absence of shot noise in this spectrum compared to the tracer autospectrum. Shot noise will explicitly decorrelate the matter density field from that of tracers at small scales. At $k \sim 0.01 \ihmpc$, we find an average reduction in uncertainty that is of an order of magnitude. This is is equivalent to averaging $N=100$ simulations at this volume. \par
We also consider variance reduction in biased tracer spectra for a slightly more accurate control variate. Instead of using the matter power spectrum, we use our Zel'dovich component fields to create linearly biased surrogates with power spectra given by
\begin{align}
    &P_{gg}^{\rm ZA} (k) = P_{mm}^{\rm ZA} (k) + 2 b_1 P_{1 \delta}^{\rm ZA} (k) + b_1^2 P_{\delta \delta}^{\rm ZA} (k), \\
    &P_{gm}^{\rm ZA} (k) = P_{mm}^{\rm ZA} (k) + b_1 P_{1 \delta}^{\rm ZA} (k). 
\end{align}
We also use $\hat{\beta}$ constructed appropriately from the disconnected approximation of Eqn.~\ref{eqn:betahat}. This step requires the re-estimation of the cross-spectra between the biased tracer and the Zel'dovich fields, and thus incurs some additional computational overhead. However this overhead is quite small, as measurements of power spectra are not a particularly computationally intensive task. For every box we estimate a value of $b_1$ using the field-level bias estimator of ref.~\cite{Kokron:2021faa} at $k_{\rm max} = 0.2 \ihmpc$, but the variance reduction obtained is insensitive to the exact value of $b_1$ adopted as long as the analytic and grid-based ZA predictions use consistent values\footnote{Indeed, the $P_{mm}$ estimator can be thought of as a special case of the linearly biased tracer example where we set $b_1 = 0$. This is an `infinitely incorrect' estimate of $b_1$, and yet we still obtain strong variance reduction.}. The results from applying our linearly biased control variate are shown in both panels of Fig.~\ref{fig:tracer_reduce}. In the left panel, we compare a single realization to the mean of the 100 \quijote HR boxes for this tracer sample and find entirely compatible results. In addition, we see that our largest scale points are wholly in agreement with predictions from the Zel'dovich approximation and our inferred linear bias value. This means that we can extend, as a model, our simulation-based measurement of $P_{g[g,m]}$ to arbitrarily large scales without worrying about sample variance. In addition, in the right panel of Fig.~\ref{fig:tracer_reduce} we show, in solid lines, the reduction in sample variance from the linearly biased control variate. The performance is mildly better than the $P_{mm}$ case, but overall comparable. The peak reduction in uncertainty over the $P_{mm}$ case is of the order of 30\% (equivalent to $\sim$70\% more volume), and occurs at $k=0.1 \ihmpc$.

\subsection{Variance reduction for hybrid EFT spectra}
\begin{figure}
    \centering
    \includegraphics[width=\textwidth]{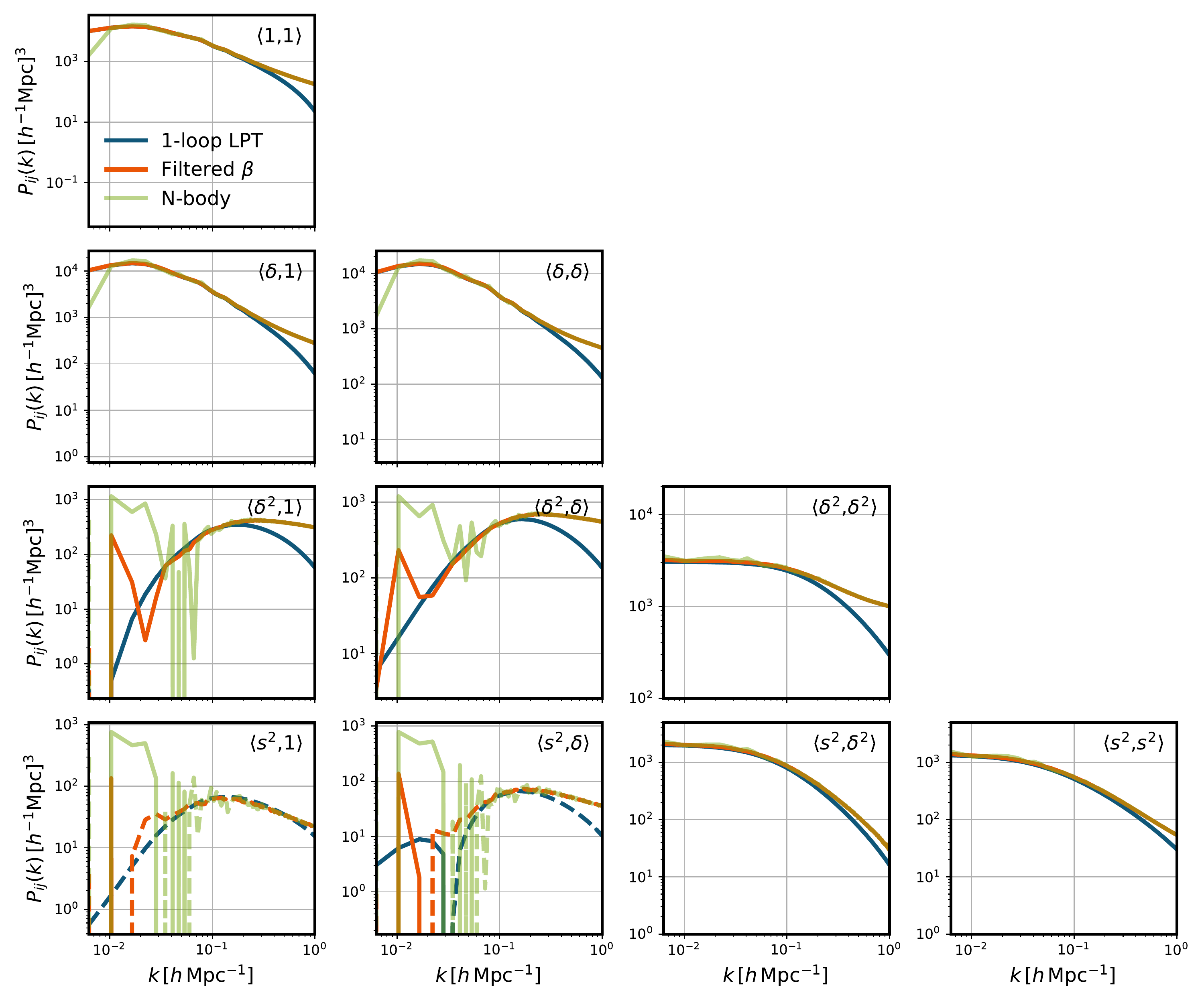}
    \caption{Comparison between the non-linear component spectra and the variance-reduced spectra measured with Zel'dovich control variates. With a single simulation we can extend accurate simulation-based measurements for the basis spectra from $k\simeq 0.1 \ihmpc$ down to $k \approx 0.03 \ihmpc$, allowing accurate matching to LPT for the extrapolation to $k\to 0$.
    }
    \label{fig:cvariables}
\end{figure}
Beyond the matter density field, or linearly biased tracers, it is also of great interest to improve measurements of all possible spectra that can contribute to the statistics of biased tracers. The question of applying control variates to the full set of basis spectra $P_{ij}(k)$ is particularly relevant for the construction of emulators of basis spectra in hybrid EFT, as discussed in \S~\ref{subsec:gridlpt}. In this subsection we report the results of applying ZA control variates to the ten basis spectra of second-order hybrid EFT. \par 
In Fig.~\ref{fig:cvariables} we show component spectra from an $N$-body simulation drawn from this suite, and the result of applying our Zel'dovich control variate scheme, as well as predictions from one-loop LPT computed using \velocileptors. We observe significant reduction in noise at large scales, especially for the cubic spectra, which enforces significantly better agreement with perturbation theory at large scales than naively obtained from a single realization. Notably, we find that a single paired Zel'dovich realization can extend the range of agreement with LPT from $k\simeq 0.1 \ihmpc$ to $k\approx 0.03 \ihmpc$. We also note that despite using ZA as the control variate, the final estimator agrees (in the $k\to 0$ limit) with one-loop Lagrangian Perturbation Theory which is a more accurate model of basis spectra. \par 
\begin{figure}
    \centering
    \includegraphics[width=\textwidth]{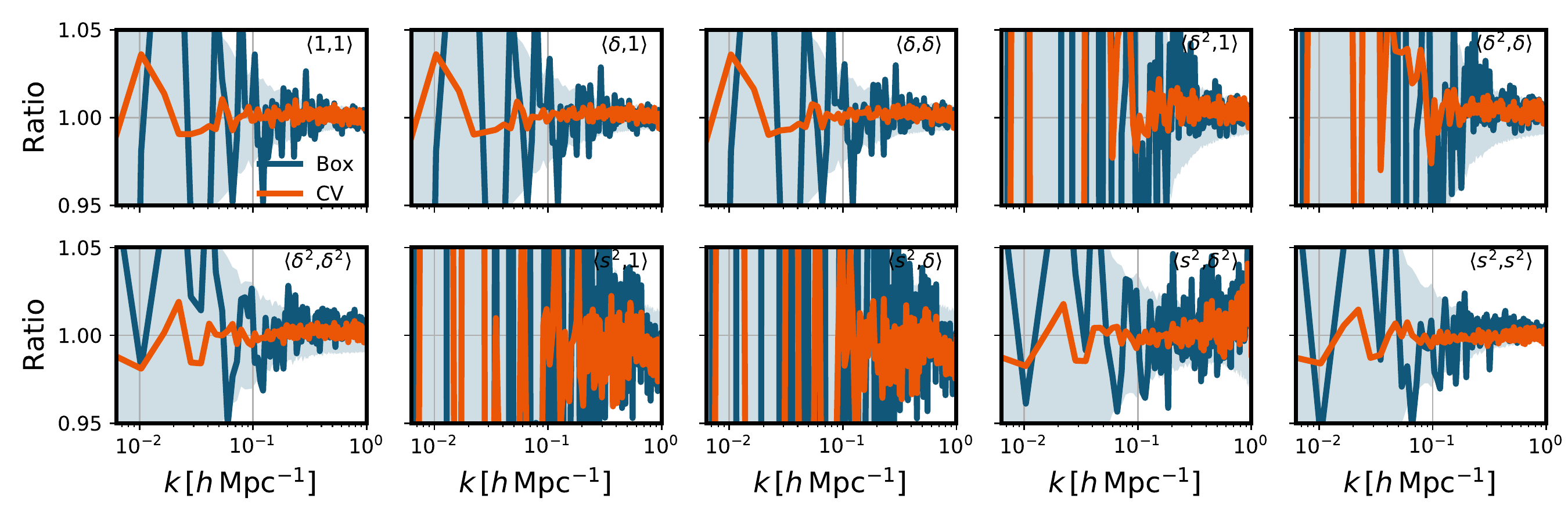}
    \caption{Ratio between basis spectra from an $N$-body box with the mean of $N=100$ \quijote HR basis spectra (in blue) and one Zel'dovich control variate estimate (in orange), using the filter of Eqn.~\ref{eqn:tanhfilter} fit to $\beta^\star_{11}$. We see significant improvements in the accuracy of the estimation of basis spectra, especially at large scales. We caution that noise in the ratios for \emph{cubic} spectra are due to the fact that even with 100 realizations the mean values of cubic spectra exhibit significant sample variance.}
    \label{fig:cvresids}
\end{figure}
A more quantitative assessment of the improvement in accuracy from our approach is shown in Fig.~\ref{fig:cvresids}, where we now show residuals of our CV approach compared to the mean of our \quijote HR boxes. We see that for non-cubic spectra the paired curves are in exquisite agreement with the low $k$ values inferred by our large volumes, with no discernible biases. For cubic spectra, this comparison is hampered by significant sample variance despite having $N=100$ boxes. The means are not well determined, and of comparable noise to a single ZA-paired control variate estimate. Since ZA realizations are extremely inexpensive to generate for a given box (and, indeed, the density field that seeds ZA displacements is a standard output of initial conditions codes), there is a large potential for improvement of the accuracy of simulation-based measurements with little extra effort relative to running the initial simulation. \par
\begin{figure}
    \centering
    \includegraphics[width=\textwidth]{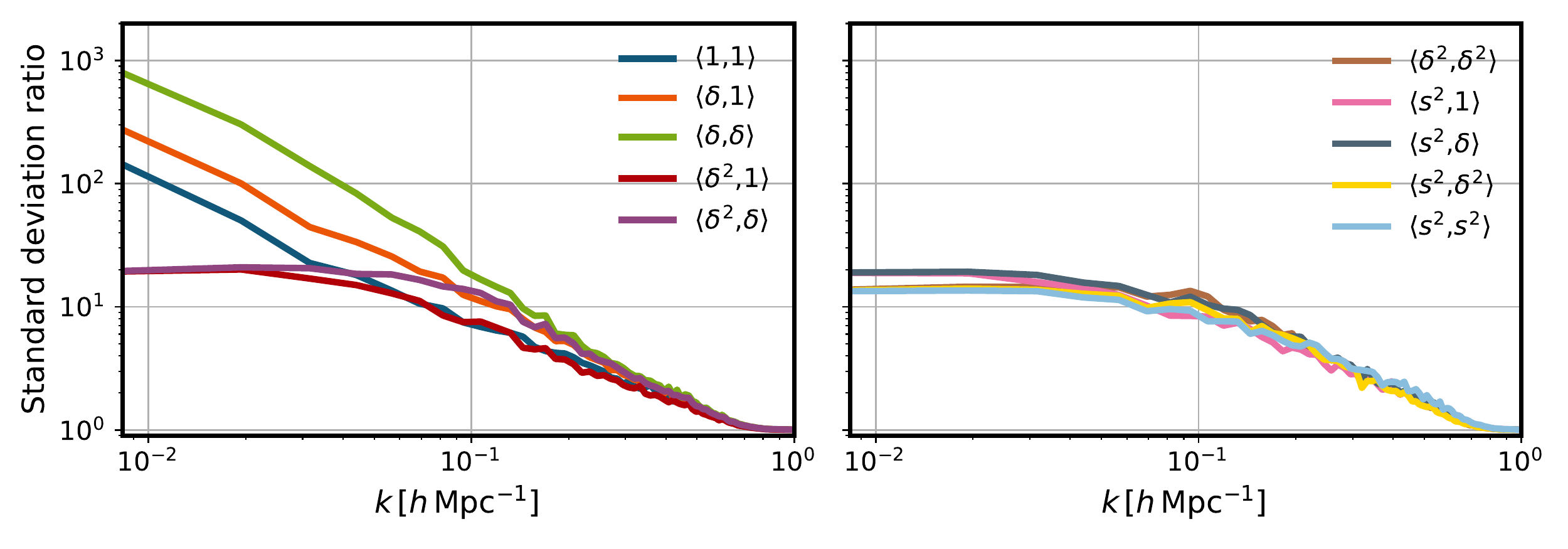}
    \caption{Ratio of the standard deviations of basis spectra, as measured from $N=100$ \quijote HR simulations, compared to the variance measured from $N=100$ pairs of simulations and ZA control variates. At large scales, we observe at least a factor of 10$\times$ reduction in uncertainty compared to cosmic variance at volumes of $V=1 ( h^{-1} {\rm Gpc})^3$ for the $\langle s^2, s^2 \rangle$ basis spectrum, up to nearly 1000$\times$ for the $\langle \delta, \delta \rangle$ spectrum. }
    \label{fig:var_reduce}
\end{figure}

  
 In Fig.~\ref{fig:var_reduce} we now show the full degree of scale-dependent variance reduction we obtain from paired  ZA control variates for all ten basis spectra that span second-order Lagrangian bias. The $\langle 1\,1\rangle$ curve was previously shown in Fig.~\ref{fig:pmm_reduce}. Quantitatively, we find that even for basis spectra of fields that couple small scale modes to large scales in the initial conditions (such as $\delta^2 (\bq)$ and $s^2 (\bq)$) we still find an over ten-fold reduction in error for all basis spectra. For spectra that don't involve convolved fields we find improvements that range from $100\times$ to $500\times$ reductions in power spectra uncertainties from our estimator which augments each simulation with a single ZA mock. \par 
The reduced performance of our variance reduction techniques for fields that involve Fourier-space convolutions can be directly understood as a consequence of the smoothing imposed in initial conditions in order to match grid-based and analytic predictions of basis spectra in the Zel'dovich approximation, as described in Appendix~\ref{appendix:calibrationsuite}. The full \quijote HR simulations have been run with un-damped initial conditions, that cut off in power at the Nyquist frequency $k_{\rm Nyq}$. When producing Lagrangian bias operators which involve Fourier-space convolutions such as
\begin{equation}
    \delta^2 (\bk) = \int \frac{d^3k'}{(2\pi)^3} \delta (\bk - \bk') \delta (\bk'),
\end{equation}
a large scale mode $\bk$ will receive contributions from arbitrarily small scales. When the linear density $\delta (\bk)$ is filtered in order to accurately match grid-based and analytic approaches, the final filtered Lagrangian fields will now be slightly de-correlated even at large scales compared to the fiducial density field used to initialize the $N$-body simulation. From Eqn.~\ref{eqn:variancered} we can then see that a lower cross-correlation coefficient will result in less substantial variance reduction than for the case of the standard density field, where the Fourier-space damping is cancelled out. This issue can be circumvented: if one produces ZA realizations from the un-damped density field whose statistics agree with analytic LPT then our technique should result in comparable variance reduction for all basis spectra. We leave a more detailed study of matching analytic and grid-based LPT to future work.\par
The consequences of applying paired ZA control variates to the problem of simulation-based modelling are manifold. The most immediate consequence is that suites of cosmological $N$-body simulations used to construct emulators such as \aemulus\cite{DeRose:2018xdj}, \texttt{BACCO} \cite{angulo2021bacco} and \texttt{AbacusSummit} \cite{Maksimova:2021ynf} can drastically improve their large-scale measurements of biased tracer spectra without running additional simulations. These methods also make clear that future suites designed for the purpose of emulating biased tracer statistics should not require prohibitively large simulation volumes to ensure precise measurements of these statistics at large scales. As long as the dynamic range of the simulation is sufficient (and most emulation suites possess better dynamic range than \quijote HR) to resolve physical scales that contribute to most basis spectra, paired ZA control variate estimators can allow for accurate emulation of large physical scales with simulations that have significantly reduced volumes. Thus, a larger number of cheaper simulations can be used to develop the suite. This larger number of simulations can then cover a larger cosmological parameter space, or perhaps known extensions to $\Lambda$CDM where the second order Lagrangian bias expansion above and Zel'dovich dynamics are still appropriate. \par 

\subsection{Comparison with previous results}
It is also worth comparing techniques described in this paper with other recent attempts at reducing the variance of basis spectra measured in simulations. In ref.~\cite{Maion22}, the authors studied the properties of basis spectra as measured in `paired-fixed' simulations. Pairing and fixing involves running, at every cosmology, two sets of simulations with non-Gaussian initial conditions that exactly fix the amplitude of fluctuations to follow $|\delta_\bk| = \sqrt{V P(k)}$ and have Fourier phases shifted by a factor of $\pi$. The authors of ref.~\cite{Maion22} have shown that by using an estimator that averages basis spectra as measured by these two sets of simulations, significant reduction in uncertainties can be obtained for a sub-set of basis spectra.\par
Our results in Fig.~\ref{fig:var_reduce} are numerically comparable to those of fig.~3 in ref.~\cite{Maion22}, however we additionally find substantial reduction in uncertainty for the quartic spectra $\langle \delta^2\, \delta^2\rangle$, $\langle \delta^2\, s^2\rangle$, and $\langle s^2\, s^2\rangle$ unlike ref.~\cite{Maion22} who find no improvement from fixing-and-pairing for those spectra. Additionally, the use of paired ZA control variates only requires cheap surrogate realizations for every simulation, with generic initial conditions, as opposed to having to run two sets of simulations with non-Gaussian initial conditions when using paired-and-fixed simulations. However, the tools of ref.~\cite{Maion22} are complementary to what we have introduced in this work and future work could investigate jointly using many variance reduction techniques to achieve greater results than what could be accomplished by each individually. \par 
Recently, the DESI collaboration \cite{Aghamousa:2016zmz} applied control variates in order to expand the volumes of their fiducial suite of simulations, \texttt{AbacusSummit}. Their control variate of choice was the approximate $N$-body solver \texttt{FastPM} \cite{Feng2016}. Over 500 surrogates were produced throughout their work, generating on the order of 400TB of ancillary data at a computational cost of 24 million NERSC CPU-hours. Their technique results in an improvement of effective volume for the redshift-space clustering of halos on the order of $100\times$. While not an apples-to-apples comparison, our worst-case scenario improvement for biased spectra is comparable to their quoted improvement. In this publication we have not analyzed redshift space correlations or higher order correlations, but we shall present these developments in future work.  \par 
The computational expense of implementing ZA control variates in an $N$-body simulation are substantially more modest than when using an approximate solver such as \texttt{FastPM}. For the code that we have developed in this paper, at the \quijote HR resolution, producing all relevant Lagrangian component fields, advecting them, and measuring all relevant cross-spectra takes on the order of 50 CPU-hours per snapshot. The largest computational expenses are equally distributed among generating the Lagrangian fields, advecting them to produce late-time realizations, and measuring all relevant cross-spectra to compute $\beta$ (which also contains all $N$-body and ZA basis spectra). \par

\section{Conclusions}
\label{sec:conclusions}

In this paper we have re-visited the problem of sample variance reduction in $N$-body simulations through use of the method of control variates. While previous works in this direction have used approximate $N$-body solvers as surrogates for structure formation, we proposed the use of the Zel'dovich approximation as a surrogate of structure formation that is simultaneously inexpensive to produce and highly correlated with the non-linear density field. We have shown that, for biased tracers in real space, we can compute the mean prediction in Zel'dovich to arbitrary precision. This development sidesteps one of the main limitations of applying control variates previously, where many realizations of the approximate simulation had to be produced in order to reduce the uncertainty on the mean variate. Additionally, we have presented a physically motivated, analytic approximation for the regression matrix $\beta$, which is often another source of significant computational expense when applying control variates. We validated this approximation using an ensemble of $N$-body simulations and showed that a simple empirical damping function applied to the approximation leads to strong agreement with numerical estimates of the regression matrix. \par 
We proceed to quantify any residual biases in our technique, as well as reduction in uncertainty, in three different problems which are applicable to many different classes of $N$-body simulations. We showed that for the matter power spectrum, we find a reduction in variance that is equivalent to averaging $10^4$ simulations at $k=0.01 \ihmpc$ for the \quijote volume. We also find strong agreement with the Zel'dovich approximation at large scales, which enables reconstructions of the configuration--space correlation function that are significantly less noisy at large scales. \par 
We then turned to the more complicated case of reducing variance in measurements of tracer statistics. We showed that using the matter power spectrum as the control variate for tracer statistics still leads to substantial variance reduction, comparable to increasing the volume of the simulation by 100$\times$ at large scales. We further showed that using a more elaborate but still highly approximate surrogate, a linearly biased tracer, led to additional reductions in uncertainty on the order of 30\% at little additional computational cost. \par 
We concluded with the most general case of applying control variates to the ten basis spectra of second-order hybrid EFT. We found improvements in effective volume that were equivalent to $\sim 100-10^6 \times$, depending on the specific basis spectrum under consideration. This dramatic increase in effective volume has substantial implications in the design and accuracy of emulators of cosmic structure. \par 
There are many future directions to consider stemming from this work. The most immediate, perhaps, is to extend our results to redshift-space. Redshift-space distortions in the Zel'dovich approximation are a simple additive offset in the original displacements. Thus, we should expect that the degree of variance reduction obtained here should be sustained. We shall present this development in a future paper. Another direction in which to extend this work is in pushing to higher order correlations such as the bispectrum of biased tracers. While the 3-point correlation function has previously been computed in Zel'dovich \cite{Tassev_2014}, the bispectrum has not. However, even lacking a full description of the bispectrum of biased tracers, the results using matter power spectra as a surrogate show that gains could be made in this regime by computing the matter bispectrum in the Zel'dovich approximation, $B_{mmm}^{\rm ZA}$.  Another potential consideration is extending these techniques to the case of simulations with primordial non-Gaussianity (PNG). Matching phases between simulations with and without PNG has recently been demonstrated to be a powerful tool to isolate the effects of PNG in simulations with substantially less sample variance \cite{Avila:2022pzw}. As biasing is well-understood in the presence of PNG \cite{Assassi_2015}, producing variance-reduced realizations of observables in the presence of PNG is well within the scope of the techniques we have laid out in this paper. We leave these, and other extensions, to future work. 
\acknowledgments
We thank Diogo Bragança and Philip Mansfield for helpful conversations. We
thank the referee for insightful comments which helped improve the
paper. N.K.~is supported by the Gerald J. Lieberman Fellowship.
J.D.~is supported by the Lawrence Berkeley National Laboratory Chamberlain Fellowship.
M.W.~and S.C.~are supported by the DOE and the NSF.
We acknowledge the use of the \quijote simulations \cite{Villaescusa_Navarro_2020} and thank their authors for making these products public.
This research has made use of NASA's Astrophysics Data System and the arXiv preprint server.
This research is supported by the Director, Office of Science, Office of High Energy Physics of the U.S. Department of Energy under Contract No. DE-AC02-05CH11231, and by the National Energy Research Scientific Computing Center, a DOE Office of Science User Facility under the same contract. Some of the computing for this project was performed on the Sherlock cluster at Stanford. We would like to thank Stanford University and the Stanford Research Computing Center for providing computational resources and support that contributed to these research results. Calculations and figures in this work have been made using \texttt{nbodykit} \citep{Hand_2018} and the SciPy Stack \citep{2020NumPy-Array,2020SciPy-NMeth,4160265}.

\appendix

\section{Basis spectra in the Zel'dovich approximation}
\label{appendix:beyondloop}

In this section we give the power spectra of tracers with bias up to quadratic order within the Zeldovich approximation. Doing so requires including terms beyond 1-loop order which are usually dropped in perturbation-theory calculations to consistently track dynamics and biasing to the same order; here we must include them to properly compute the mean of the basis spectra in simulations where the dynamics are taken to be Zeldovich. Up to 1-loop order the results below are equivalent to those in Eqn.~4.11 from ref.~\cite{Chen_2020} with terms due to beyond-Zeldovich displacements set to zero, while a subset of the higher-order terms for the case of quadratic density bias were calculated in ref.~\cite{White:2014gfa,2015MNRAS.447.2169W}.

To evaluate the basis power spectra within the Zeldovich approximation we need to evaluate the functional
\begin{equation*}
    \mathcal{Z}(\bq, \lambda_n, a_{n}) = \langle e^{\mathcal{M}(\bq_1, \bq_2, \lambda_n, a_{n})} \rangle
\end{equation*}
where the exponent is defined as:
\begin{equation}
\label{eqn:exponent}
    \mathcal{M}(\bq_1, \bq_2) = i \bk \cdot \Delta + \sum_{n=1,2} \lambda_n \delta(\bq_n) + a_{n,ij} s_{ij}(\bq_n), \quad \Delta = \Psi(\bq_1) - \Psi(\bq_2).
\end{equation}
Since $\mathcal{M}$ is Gaussian we have that $\mathcal{Z}$ is simply given by the exponentiated second cumulant,
\begin{align*}
    \frac{1}{2} \langle \mathcal{M}^2 \rangle_c = &- \frac{1}{2} k_i k_j A_{ij} + (\lambda_1 + \lambda_2) i k_i U_i + \lambda_1 \lambda_2 \xi_L + \frac{1}{2}(\lambda_1^2 + \lambda_2^2) \sigma_L^2 \\
    &+  a_{1,ij} a_{2,kl} C_{ijkl} + \frac{1}{2}(a_{1,ij}a_{1,kl} + a_{2,ij}a_{2,kl}) \langle s_{ij} s_{kl} \rangle \\
    &+ (a_{1,jk} + a_{2,jk}) i k_i B_{ijk} +  (\lambda_1 a_{2,ij} + \lambda_2 a_{1,ij}) E_{ij} 
\end{align*}
where we have defined the functions \cite{Carlson_2012}
\begin{equation}
    A_{ij}(\bq) = \langle \Delta_i \Delta_j \rangle = X(q) \delta_{ij} + Y(q) \hat{q}_i \hat{q}_j \quad , \quad U_i(\bq) = \langle \delta_1 \Delta_i \rangle
\end{equation}
and we also define shear correlators
\begin{equation}
    C_{ijkl} = \langle s_{1,ij} s_{2,kl} \rangle \quad , \quad
    B_{ijk} = \langle \Delta_i s_{1,jk} \rangle \quad , \quad
    E_{ij} = \langle \delta_1 s_{2,ij} \rangle
\end{equation}
and we have used that $\langle \delta s_{ij} \rangle = 0$. All the correlators are strictly functions of $\bq = \bq_1 - \bq_2$ by translation invariance.

In order to obtain the biased-tracer power spectrum we use the substitutions
\begin{equation}
    b_1 \delta_n \rightarrow b_1 \frac{d}{d\lambda_n} , \quad
    b_2 \delta_n^2 \rightarrow b_2 \frac{d^2}{d\lambda_n^2} \quad , \quad
    b_s s^2_n \rightarrow b_s \delta_{ia} \delta_{jb}\ \frac{d}{d a_{n,ij}} \frac{d}{d a_{n,ab}}.
\end{equation}
in the bias functional $F(\bq_n)$, which becomes an operator $\hat{F}_n$. The power spectrum is then given by
\begin{equation}
    P(k) = \int d^3\bq\ e^{i\bk \cdot \bq}
    \Big[\hat{F}_1 \hat{F}_2 \mathcal{Z}(\bq, \lambda_n, a_{n}) \Big]_{\lambda_n, a_n=0}
    \quad .
\end{equation}
This gives the component spectra as
\begin{equation}
    P_{ij}(k) = \int d^3\bq\ e^{i\bk \cdot \bq-\frac{1}{2}k_i k_j A_{ij}}\ F_{ij}(\bk,\bq)
\label{eqn:componentspectrum}
\end{equation}
where for each basis spectrum the function $F_{ij}(\bk,\bq)$ is given by
\begin{align}
\label{eqn:basiskernels}
    (1,1) &: 1 \nonumber \\
    (1,b_1) &:  i k_i U_i\nonumber \\
    (b_1,b_1) &: \xi_L - k_i k_j U_i U_j \nonumber \\
    (1, b_2) &: - k_i k_j U_i U_j \nonumber\\
    (b_1, b_2) &: 2 i k_i U_i \xi_L - i k_i k_j k_k U_i U_j U_k \nonumber \\
    (b_2, b_2) &: 2\xi_L^2 - 4 k_i k_j U_i U_j \xi_L + k_i k_j k_k k_l U_i U_j U_k U_l \nonumber \\
    (1,b_s) &: - k_i k_j B_{iab} B_{jab} \nonumber \\
    (b_1, b_s) &: 2 i k_i B_{iab} E_{ab} - i k_i k_j k_k U_i B_{jab} B_{kab} \nonumber \\
    (b_2, b_s) &: 2 E_{ab} E_{ab} - 4 k_i k_j U_i E_{ab} B_{jab} + k_i k_j k_k k_l U_i U_j B_{kab} B_{lab} \nonumber \\
    (b_s, b_s) &:2 C_{abcd} C_{abcd} - 4 k_i k_j B_{iab} B_{jcd} C_{abcd} + k_i k_j k_k k_l B_{iab} B_{jab} B_{kcd} B_{lcd}.
\end{align}

Many of the contractions in the above can be written as correlators of the scalar $s^2$ with densities and displacements. Some of these have been previously defined in other works (see e.g.\ ref.~\cite{Chen_2021})
\begin{align}
    \chi = \langle \delta^2_1 s^2_2\rangle_c = 2 E_{ab} E_{ab} \quad ,
    & \quad\zeta = \langle s_1^2 s_2^2 \rangle_c = 2 C_{abcd} C_{abcd} \nonumber \\
    \Upsilon_{ij} = \langle s^2 \Delta_i \Delta_j \rangle = 2 B_{iab} B_{jab} \quad ,
    & \quad V^{12}_i = \langle \delta_1 s^2_2 \Delta_i \rangle = 2 E_{ab} B_{iab}
\end{align}
so we can simplify some of above expressions for basis spectra:
\begin{align}
    (1, b_s) &: - \frac{1}{2} k_i k_j \Upsilon_{ij} \nonumber \\
    (b_1, b_s) &: i k_i V^{12}_i - \frac{1}{2} i k_i k_j k_k U_i \Upsilon_{jk} \nonumber \\
    (b_2, b_s) &: \chi - 2 k_i k_j U_i V^{12}_j + \frac{1}{2} k_i k_j k_k k_l U_i U_j \Upsilon_{kl} \nonumber \\
    (b_s, b_s) &: \zeta - 4 k_i k_j B_{iab} B_{jcd} C_{abcd} + \frac{1}{4} k_i k_j k_k k_l \Upsilon_{ij} \Upsilon_{kl}.
\end{align}

The $(b_s,b_s)$ component contains a term that cannot be reduced to previously computed quantities but is rather proportional to $L_{ij} = B_{iab} B_{jcd} C_{abcd}$. This term comes from the expectation value of $(1/2)(i k_i \Delta_i)^2 s_1^2 s_2^2$ which contains
\begin{equation*}
    \langle \Delta_i \Delta_j s_1^2 s_2^2 \rangle \ni 8 \langle \Delta_i s_{1,ab} \rangle \langle \Delta_j s_{2,cd} \rangle \langle s_{1,ab} s_{2,cd}\rangle
\end{equation*}
leading to the combinatorial factor $(-4) = \frac{1}{2} \times i^2 \times 8$. This contribution can be expressed in terms of its components as
\begin{equation}
    L_{ij} = 2 K_1 J_3 \delta_{ij} + \big( 2K_1 (J_2+J_3+J_4) + K_2 (J_2+2J_3+J_4) \big) \hat{q}_i \hat{q}_j
\end{equation}
where we the $J_n$ are defined as in Appendix A of ref.~\cite{White:2014gfa,2015MNRAS.447.2169W} and we have defined the quantities
\begin{equation}
    K_1 = 2 J_3(J_6+J_8), \quad K_2 = 2 J_3 (J_7 +2J_8 + J_9) + J_4 (2J_6+J_7+4J_8+J_9).
\end{equation}
All of the above expressions are implemented in the publicly available code \zenbu\footnote{Ze(ldovich calculations for) N-B(ody Em)u(lators); \url{https://github.com/sfschen/ZeNBu}.}.

\section{Matching analytic and grid-based LPT}
\label{appendix:calibrationsuite}

\begin{figure}
    \centering
    \includegraphics[width=\textwidth]{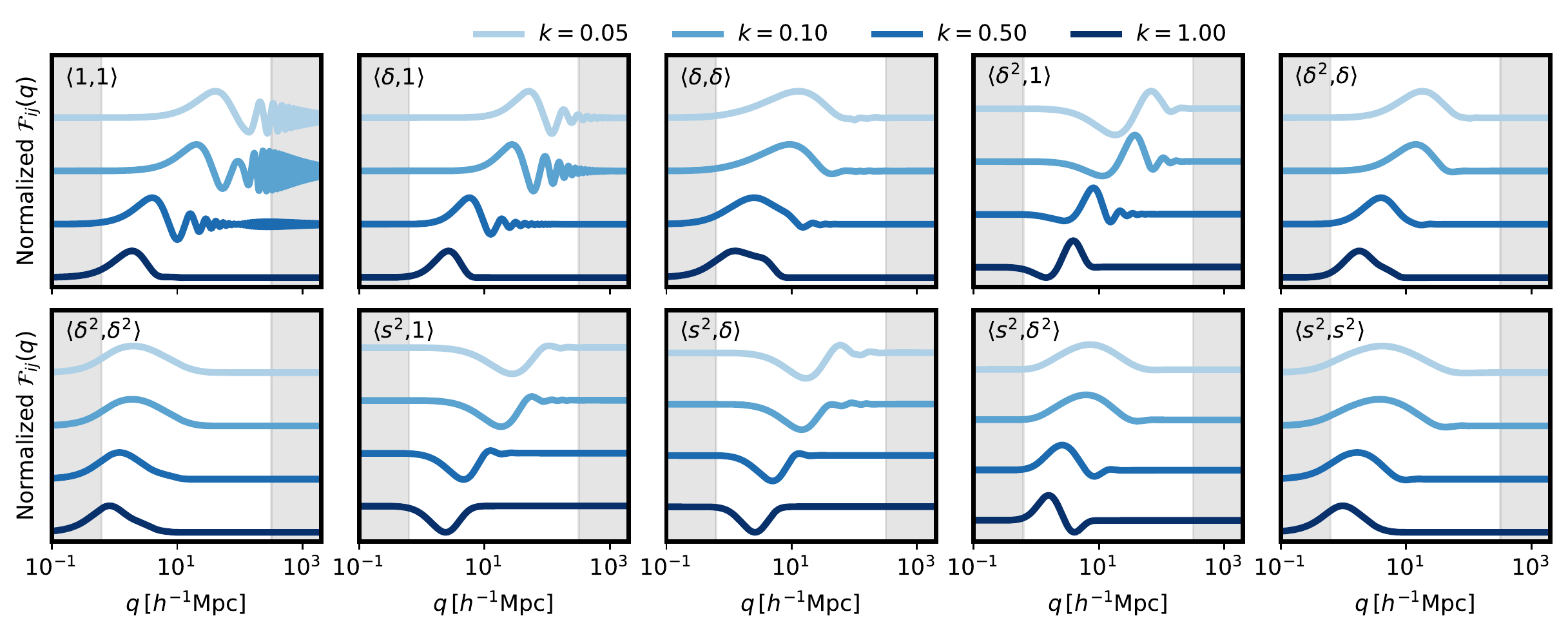}
    \caption{Normalized integrands for Lagrangian basis spectra at scales of $k = [0.05, 0.1, 0.5, 1] \ihmpc$. The shaded regions correspond to physical scales that are not well probed by the fiducial \quijote HR simulations. Each integrand is offset (vertically) by an arbitrary value so their features and support can be clearly distinguished. }
    \label{fig:integrands}
\end{figure}

The success of the control variates approach we have laid out in this work is predicated on ensuring precise agreement from our two separate predictions for basis spectra in the Zel'dovich approximation. If grid-based calculations do not agree with the analytic result, then the control variate could introduce significant biases, as well as additional uncertainty, in the estimates of the fully non-linear spectra. \par 
In this appendix we explore, in more detail, the structure of the contributions to any given basis spectrum in the analytic code. Any given spectrum $P_{ij} (k)$ is written as in the form of Eqn.~\ref{eqn:componentspectrum}. In order to efficiently evaluate the integrals, \zenbu\  further projects the $F_{ij}(\bk, \bq)$ kernels into a Spherical Fourier-Bessel basis. The angular integrals are then performed analytically, and the radial integrals reduce to Hankel transforms of the form
\begin{equation}
\label{eqn:hankel}
    P_{ij} (k) = 4\pi \sum_{\ell=0}^\infty \int q^2 dq\ e^{-k^2 (X + Y)/2} \left (\frac{kY}{q}\right)^\ell f_{ij}^{(\ell)}(q,k) j_\ell (kq),
\end{equation}
where $f_{ij}^{(\ell)}(q)$ is a basis spectrum-dependent kernel. These Hankel transforms can then be rapidly evaluated using the FFTLog algorithm \cite{Hamilton_2000,Vlah15,Schmittfull_2016}. A summary of the structure of the implementation can be found in Appendix A of ref.~\cite{white2022cosmological}, with full details available in refs.~\cite{Vlah_2016,Chen_2020}. In practice, additional regularization is needed to compute this expression stably and this is done by including a cutoff in the linear power spectrum
\begin{equation}
\label{eqn:cutoff}
    P^{\rm lin} (k) \to e^{-(k/k_{\rm cut})^2} P^{\rm lin} (k), 
\end{equation}
where $k_{\rm cut}$ is a cutoff scale imposed from filtering the linear density modes at initial conditions as
\begin{equation}
    \delta (\bk) \to e^{-(k/k_{\rm cut})^2/2} \delta(\bk) .
\end{equation}
This same filter is applied to the initial density field used to evaluate Zel'dovich displacements and their subsequent component fields/spectra. We use $k_{\rm cut} = \pi N_{\rm p}^{1/3} / L_{\rm box} \approx 3.2 \ihmpc$,  where $N_p$ is the total number of particles in the simulation, corresponding to the smallest Fourier modes probed by the initial conditions of the \quijote HR boxes\footnote{As \quijote is evolved using a tree-based code, nonlinear forces are calculated at scales smaller than $k_{\rm cut}$.}. Without this equivalent filtering, we find that we are unable to match grid-based results and analytics, even in the case of $P_{mm}(k)$.\par 
If we re-cast Eq.~\ref{eqn:hankel} in the following form:
\begin{equation}
    P_{ij}(k) = \int dq\ \mathcal{F}_{ij} (q,k),
\end{equation}
where $\mathcal{F}_{ij}$ is given by
\begin{equation}
\label{eqn:configkernel}
    \mathcal{F}_{ij} (q,k) =  \sum_\ell  e^{-(k^2/2) (X^{\rm lin} + Y^{\rm lin})} \left ( \frac{kY}{q} \right )^\ell f_{ij}^{(\ell)} (q,k) q^2j_\ell (kq),
\end{equation}
then we can assess the scales needed to be probed by our grid-based realization in order to ensure accuracy relative to the analytic calculation. If for a given wavenumber $k$, $\mathcal{F}_{ij}(q,k)$ has significant support for $q$ that are either larger than the largest resolved mode in the box, $q_{\rm max}$ or smaller than the smallest scale $q_{\rm min}$, then these contributions will be missed in the grid-based calculation. While at first impression one might assume that $q_{\rm min} = L_{\rm box}/N_{\rm mesh}$, the fundamental grid spacing,  and $q_{\rm max} = L_{\rm box}/2$, the largest resolvable separation, this is not the case. Power in the $N$-body simulation is initialized on a Fourier-space grid, which is then Fourier-transformed to configuration space. This imposes two anisotropic window functions which filter small and larger scales. The ``true'' density field is thus 
\begin{equation}
    \tilde{\delta}(\bk) = W_{\rm  UV} (\bk) W_{\rm  IR}(\bk) \delta(\bk),
\end{equation}
where the small-scale and large-scale anisotropic window functions are, respectively,
\begin{equation}
    W_{\rm  UV} (\bk) = \begin{cases}
    1,\quad {\rm if } \quad |k_i| \leq \frac{\pi N_{\rm mesh}}{L_{\rm box}} \\
    0,\quad {\rm if } \quad |k_i| > \frac{\pi N_{\rm mesh}}{L_{\rm box}} 
    \end{cases}, \quad \quad   W_{\rm  IR} (\bk) = \begin{cases}
    1,\quad {\rm if } \quad |k_i| \geq \frac{2\pi }{L_{\rm box}} \\
    0,\quad {\rm if } \quad |k_i| < \frac{2\pi}{L_{\rm box}}
    \end{cases}.
\end{equation}
The sharp Fourier-space windows imposed by the gridded density field will leak power into scales $q < L_{\rm box}/N_{\rm mesh}$.  In configuration space, these filters are given by sinc functions of the form
\begin{equation}
    W_{\rm UV} (\bq) \propto \prod_i {\rm sinc} \left ( \frac{\pi N_{\rm mesh} q_i}{L_{\rm box}} \right ).
\end{equation}
If we compare the integral of this window relative to a `sharp-$q$' window function, we can define an equivalent $q_{\rm min}$ from assessing when their integrals reach their asymptotic values at $q\to \infty$. For the sinc function with argument as we've defined, this value is reached at roughly $q_{\rm min} \approx 2/k_{\rm Nyq}$\footnote{The precise value is closer to $1.926 / k_{\rm Nyq}$ and is related to when the Sine integral ${\rm Si}(q_{\star}) = \pi / 2$ for the first time.}. Thus, we adopt $q_{\rm min} = 2/k_{\rm Nyq}$ as a heuristic for the smallest configuration-space scales probed by simulations when understanding the extent of integrands $\mathcal{F}_{ij}(q)$. Similarly, we choose $q_{\rm max} = 2 / k_f = L_{\rm box} / \pi$ as a heuristic scale for the largest configuration-space scale. In Fig.~\ref{fig:integrands}, we explicitly visualize the kernels $\mathcal{F}_{ij}$ for $k \in [0.05, 0.1, 0.5, 1]$. We find that across this range of scales, most integrands have support well within the scales probed by the \quijote HR boxes. We also note that this same figure explains why we cannot use the fiducial resolution \quijote suite: for $k\ge 0.5 \ihmpc$ they fail to resolve scales near the peak of the integrands for several of the basis functions. However, we also note that from our heuristic $q_{\rm min}$ we might be at risk of missing several low-$q$ scales for spectra such as $\langle s^2 s^2\rangle$ and $\langle \delta^2 s^2 \rangle$. \par 
\begin{figure}
    \centering
    \includegraphics[width=\textwidth]{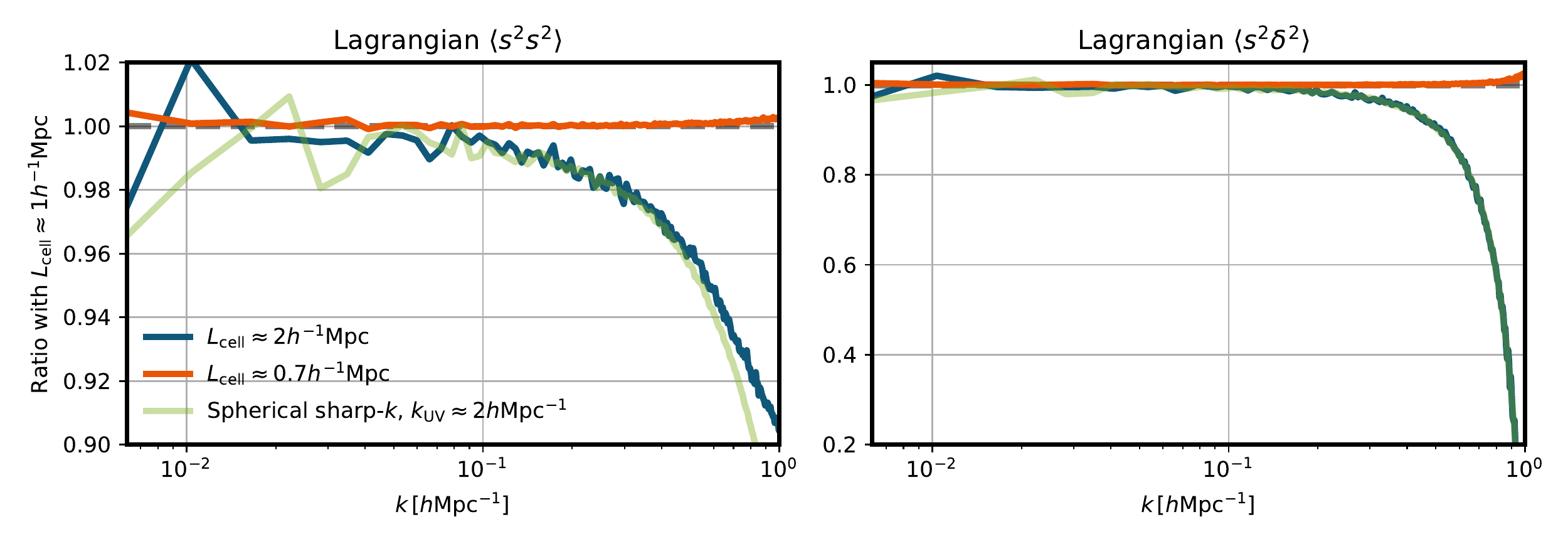}
    \caption{The resolution dependence of basis spectra at Lagrangian coordinates. \emph{Left:} The ratio of a standard \quijote resolution realization, a super-resolution realization and an analytic calculation of the effects of a small scale cutoff on the $\langle s^2 s^2 \rangle$ spectrum compared to grids computed at the fiducial \quijote HR resolution of this work. \emph{Right:} The same ratios but for the basis spectrum $\langle s^2 \delta \rangle$. Cutting off the linear theory power spectrum at small scales produces an effect on correlators that is comparable to the effect of reducing grid size.
    }
    \label{fig:lagcor}
\end{figure}
Another way to assess whether the dynamic range in a grid used for grid-based LPT is sufficient is to consider, instead, Lagrangian correlators. These are the leading contributions to the final advected spectra, but don't mix configuration-space and Fourier-space modes as strongly as in the post-advection case we have considered above. For  $\langle \delta^2 s^2\rangle$ and $\langle s^2 s^2\rangle$ these correlators are given by Hankel transforms of the functions $\chi(q)$ and $\zeta(q)$ defined in Appendix.~\ref{appendix:beyondloop}. Discrepancies in Lagrangian correlators will propagate into post-advection measurements, and their origin similarly comes from being unable to resolve all scales that contribute to a given spectrum in a grid-based approach. To illustrate this we consider the \emph{Lagrangian} power spectra
\begin{align}
\label{eqn:chihankel}
    P_{\delta^2 s^2}^{\rm Lag} (k) = 4\pi \int q^2\, dq
    \   j_0 (kq)\, \chi(q),\\
    \nonumber P_{s^2 s^2}^{\rm Lag} (k) = 4\pi \int q^2\, dq
    \   j_0(kq)\, \zeta(q).
\end{align}
We measure, in the Lagrangian fields, these basis spectra for three grids that share the same initial seeds and power spectra with varying resolutions. Beyond the fiducial \quijote HR resolution, we also consider a grid at the standard \quijote resolution of $N_{\rm mesh} = 512$. To understand whether our \quijote HR results are well converged, we also consider correlators measured at a resolution comparable to that of the \aemulus suite, using $N_{\rm mesh} = 1400$. We also compute the spectra in Eqn.~\ref{eqn:chihankel} analytically in two different ways. The first is using, as input, the linear theory damped $P(k)$ across all scales. If our grid adequately resolves all relevant physical scales, and the numerical operation of extracting $s_{ij}$ is well-converged, then the comparison between $P_{\delta^2s^2}^{\rm Lag}(k)$ computed using Eqn.~\ref{eqn:chihankel} and as measured in the Lagrangian fields of our simulations should be excellent at small scales where sample variance is negligible. We have verified that in our boxes this is the case -- residuals between grid-based and analytic Lagrangian correlators agree to within 0.2\% at $k=1 \ihmpc$. \par
Our second calculation uses analytic window functions that are \emph{spherically} sharp in $k$ space, with $k_{\rm  UV} = 512\pi  / L_{\rm box}$ and $k_{\rm  IR} = 2\pi / L_{\rm box}$. While these window functions are different from how the grid-based calculations are treated, their inclusion in the analytic calculation allows for comparison with the downsampled result at $N_{\rm mesh} = 512$. We increase $k_{\rm  UV}$ by 20\% to approximately convert the spherical damping back to the cartesian case employed in grid-based calculations, resulting in $k_{\rm UV} \approx 2 h{\rm Mpc}^{-1}$. We show these results in Fig.~\ref{fig:lagcor}, where we display the ratio of power spectra at various resolutions relative to the fiducial \quijote HR realization. The higher resolution realization has less than a percent additional power at $k=1 \ihmpc$ for $\langle s^2 s^2\rangle$ while the excess is of order a percent for $\langle s^2 \delta^2 \rangle$. This excess for $\langle s^2 \delta^2 \rangle$ can also be seen in Fig.~\ref{fig:zaratio_bin}, implying for that specific spectrum we would find potentially better convergence with the analytic ZA result at a higher resolution. The comparison with the \aemulus resolution boxes shows that we are not missing small-scale modes for the \quijote HR-based analysis. \par 
Turning to the standard \quijote resolution realizations, we find a large absence of power for both Lagrangian correlators. At $k=1 \ihmpc$ the deficit for $\langle s^2 s^2 \rangle$ is on the order of 10\%, and both this amount as well as the shape of the decay is well matched by our theoretical result smoothed by a spherical sharp-$k$ window. For $\langle s^2 \delta^2 \rangle$ the effects are even stronger, we find on the order of 80\% missing power at $k=1 \ihmpc$. Again, the observed damping is well described by convolving the input theory spectrum with a spherical sharp-$k$ window function before computing correlators in \zenbu. Fig.~\ref{fig:lagcor} shows that the standard resolution \quijote simulations do not possess sufficient dynamic range for Zel'dovich control variates to be used effectively to scales of $k=1 \ihmpc$.\par 
Future attempts to use control variates should be cognizant of the analysis carried out in this Appendix. Understanding accuracy requirements at different scales for emulation, as well as what the dynamic range required in a simulation/grid-based LPT in order to achieve satisfactory prediction of the mean of a control variate can be obtained via the arguments laid out here. 
\section{Challenges in going beyond Zel'dovich}
\label{appendix:beyondza}

In this work we have concerned ourselves solely with using the Zel'dovich approximation as the control variate that correlates with the non-linear density field. It is natural, then, to wonder whether even further reduction in sample variance can be achieved with a surrogate that is more closely correlated to the non-linear density, but is still analytic. The natural candidate for such a surrogate is the density field predicted from displacements carried out at higher order in LPT. Indeed, the state of the field is such that there now exist mature codes that compute basis spectra in one-loop LPT (such as \velocileptors) and using grid-based schemes, such as \monofonic \citep{Michaux:2020yis}. \par 
However, there is an additional subtlety if one wishes to compare 3LPT between an analytic prediction and a grid-based realization at the same order. Most analytic calculations codes for higher order LPT implement so-called \emph{convolution Lagrangian pertrubation theory} \cite{Carlson_2012}. CLPT performs a re-summation of certain terms that arise in the expansion of connected moments, $\langle \mathcal{M}^n \rangle_c$, with $\mathcal{M}$ as defined in Eqn.~\ref{eqn:exponent}. In the Zel'dovich approximation, all ingredients in the expansion are Gaussian and thus there are no connected moments beyond second order. \par 
However, when $\Delta$ contains contributions from higher order displacements the above statement is no longer true. Now, connected moments beyond the second exist and will contribute to the evaluation of $\mathcal{Z}$. For example, a contribution arises from the correlation of three displacements at second order
\begin{equation}
    W_{ijk} = \langle \Delta_i \Delta_j \Delta_k \rangle.
\end{equation}
In CLPT, and other perturbative approaches, higher order terms like $W_{ijk}$ are treated as small and expanded out of the exponential in $\mathcal{Z}$\footnote{See \cite{Vlah15_loop} for a discussion specifically about expanding $W_{ijk}$ or not. There are many terms beyond $W_{ijk}$ which are expanded from this exponential in perturbative approaches.}. On the other hand, grid-based approaches fully evaluate this exponential. These differences lead to discrepancies between grid-based and analytic LPT when trying to extend the use of surrogates to higher order. Attempts to use higher order LPT as a control variate must, then, deal with this issue before being practical. We leave this question to further work, as we find the performance of the Zel'dovich approximation as a control variate to be adequate for the purposes of a first study.

\bibliography{main}
\bibliographystyle{jhep}

\end{document}